\documentclass[preprint,11pt]{elsarticle}

\usepackage{amssymb}

\usepackage{lineno}

\biboptions{compress}

\usepackage{wasysym}
\usepackage{multirow}
\usepackage{ctable}
\usepackage{sidecap}
\usepackage{graphicx,amssymb,amstext} 
\usepackage{lscape}

\begin{document}

\newcommand{\swift}{Swift}
\newcommand{\fermi}{Fermi}
\newcommand{\xray}{\mbox{X-ray}}
\newcommand{\gray}{\mbox{$\gamma$-ray}}

\def\simlt{\mathrel{\hbox{\rlap{\hbox{\lower4pt\hbox{$\sim$}}}\hbox{$<$}}}}
\def\simgt{\mathrel{\hbox{\rlap{\hbox{\lower4pt\hbox{$\sim$}}}\hbox{$>$}}}}

\newcommand{\fix}{\textbf{???}}
\newcommand{\url}[1]{\texttt{#1}}

\def\Veff{V^{(\text{eff})}}
\def\Aeff{A^{(\text{eff})}}
\def\Vav{V^{(\text{eff})}} 
\def\rpbh{\rho_{\rm PBH}}
\def\rgwnu{\rho_{\text{GW-CN}}}

\def\nhen{n_\text{HEN}}
\def\DGW{D_{\rm GW}}
\def\DGWHEN{D_{\rm GWHEN}}
\def\EGW{E^{\text{iso}}_{\text{GW}}}
\def\EHEN{E^{\text{iso}}_{\nu}}
\def\Msun{M_{\astrosun}}
\def\peryear{year$^{-1}$}
\def\Enu{\mathcal{E}_{\nu}}
\newcommand{\Enuk}[1]{\mathcal{E}_{\nu#1}}

\newcommand\FPR[1]{R^{\text{(FP)}}_{#1}}
\newcommand\FPRname[1]{R^{\text{(FP)}}_{\text{\tiny{#1}}}}

\newcommand\TPR[1]{R^{\text{(TP)}}_{#1}}
\newcommand\TPRname[1]{R^{\text{(TP)}}_{\text{\tiny{#1}}}}

\def\HEN{CN}
\def\HENlong{cosmogenic neutrinos}
\def\HENsing{cosmogenic neutrino}

\def\refe{\par \noindent \hangindent=1pc \hangafter=1 }
\def\barnu{$\nu\bar\nu$~}
\def\mathnew{\mathsurround=0pt}
\def\simov#1#2{\lower .5pt\vbox{\baselineskip0pt \lineskip-.5pt
       \ialign{$\mathnew#1\hfil##\hfil$\crcr#2\crcr\sim\crcr}}}
\def\simg{\mathrel{\mathpalette\simov >}}
\def\siml{\mathrel{\mathpalette\simov <}}
\def\gele{\stackrel{>}{<}}
\def\lege{\stackrel{<}{>}}
\def\Mesz{M\'esz\'aros\,}
\def\Pacz{Paczy\'nski\,}
\def\Schr{Schr\"odinger\,}
\def\Sch{Schr\"odinger\,}
\def\Schw{Schwarzschild\,}
\def\U{Universe\,}
\def\glos{\glossary}
\def\ctl{\centerline}
\def\name{name=}
\def\desc{description}
\def\dx{description}
\def\lbr{\linebreak}
\def\msun{M_\odot}
\def\lsun{L_\odot}
\def\etal{{\it et~al.}}
\def\nonum{\nonumber}
\def\mfp{m.f.p.~}
\def\arcdeg{\hbox{$^\circ$}}
\def\arcmin{\hbox{$^\prime$}}
\def\arcsec{\hbox{$^{\prime\prime}$}}
\def\vareps{\varepsilon}
\def\eps{\epsilon}
\def\lrar{\leftrightarrow}
\def\Rarrow{\Rightarrow}
\def\noind{\noindent}
\def\nobu{\noindent$\bullet$~}
\def\bitm{\bibitem}
\def\para{\parallel}
\def\beq{\begin{equation}}
\def\enq{\end{equation}}
\def\bea{\begin{eqnarray}}
\def\ena{\end{eqnarray}}
\def\bec{\begin{center}}
\def\enc{\end{center}}
\def\bsk{\bigskip}
\def\msk{\mediumskip}
\def\ssk{\smallskip}
\def\noi{\noindent}
\def\bull{\bullet}
\def\blist{\begin{list}{$\bullet$}{\itemsep 0.0in \parsep 0.0in}}
\def\elist{\end{list}}
\def\bitem{\begin{list}{\arabic{enumi}.}{\usecounter{enumi} \itemsep 0.0in \parsep 0.0in}}
\def\eitem{\end{list}}
\def\cm{\hbox{~cm}}
\def\s{\hbox{~s}}
\def\erg{\hbox{~erg}}
\def\g{\hbox{~g}}
\def\dyne{\hbox{~dyne}}
\def\fm{\hbox{~fm}}
\def\b{{\rm b}}
\def\mb{{\rm mb}}
\def\mub{{\mu{\rm b}}}
\def\nb{{\rm nb}}
\def\pb{{\rm pb}}
\def\EeV{\hbox{EeV}}
\def\PeV{\hbox{PeV}}
\def\TeV{\hbox{~TeV}}
\def\GeV{\hbox{~GeV}}
\def\MeV{\hbox{~MeV}}
\def\keV{\hbox{~keV}}
\def\eV{\hbox{~eV}}
\def\deg{\hbox{~deg}}
\def\K{{~\hbox{K}}}
\def\Hz{\hbox{~Hz}}
\def\gmunu{g_{\mu\nu}}
\def\etamunu{\eta_{\mu\nu}}
\def\Rmunu{R_{\mu\nu}}
\def\Tmunu{T_{\mu\nu}}
\def\munu{{\mu\nu}}
\def\albet{{\alpha\beta}}
\def\pv{{\vec p}}
\def\rv{{\vec r}}
\def\vv{{\vec v}}
\def\vv{{\vec v}}
\def\vbet{{\vec \beta}}
\def\vnab{{\vec \nabla}}
\def\xv{{\vec x}}
\def\kv{{\vec k}}
\def\qv{{\vec q}}
\def\Bv{{\vec B}}
\def\Ev{{\vec E}}
\def\xp{{x'}}
\def\rph{r_{ph}}
\def\rphem{r_{ph,\gamma}}
\def\rphpi{r_{ph,\pi}}
\def\phot{photosphere~}
\def\hgamma{{\hat \gamma}}
\def\hatgam{{\hat \gamma}}
\def\gamu{\gamma^\mu}
\def\part{\partial}
\def\bpsi{{\boldsymbol\psi}}
\def\barbpsi{{\bar{\boldsymbol\psi}}}
\def\blambda{{\boldsymbol\lambda}}
\def\bsigma{{\boldsymbol\sigma}}
\def\barnu{{\bar \nu}}
\def\barpsi{{\bar \psi}}
\def\barX{{\bar X}}
\def\barp{{\bar p}}
\def\barq{{\bar q}}
\def\barQ{{\bar Q}}
\def\barq{{\bar q}}
\def\barell{{\bar \ell}}
\def\nue{\nu_e}
\def\barnue{{\bar \nu_e}} 
\def\numu{\nu_\mu}
\def\barnumu{{\bar \nu}_\mu}
\def\nutau{\nu_\tau}
\def\barnutau{{\bar \nu}_\tau}
\def\nul{\nu_L}
\def\nur{\nu_R}
\def\er{e_R}
\def\el{e_L}
\def\psil{\psi_L}
\def\psir{\psi_R}
\def\km{{\rm km}}
\def\kg{{\rm kg}}
\def\pc{{\rm pc}}
\def\kpc{{\rm kpc}}
\def\Mpc{{\rm Mpc}}
\def\Gpc{{\rm Gpc}}
\def\sr{{\rm sr}}
\def\hr{{\rm hr}}
\def\yr{{\rm yr}}
\def\Gyr{{\rm Gyr}}
\def\Tgut{T_{GUT}}
\def\Tew{T_{EW}}
\def\mpl{m_{Pl}}
\def\Tpl{T_{Pl}}
\def\h75{h_{75}}
\def\Omh75{\Omega h^2_{75}}
\def\aeq{a_{eq}}
\def\adec{a_{dec}}
\def\bCP{{\bf CP}}
\def\bC{{\bf C}}
\def\bP{{\bf P}}
\def\bCPT{{\bf CPT}}
\def\bB{{\bf B}}
\def\bL{{\bf L}}
\def\bT{{\rm T}}
\def\bs{\boldmath}
\def\bsy{\boldsymbol}
\def\bp{\bs{p}}
\def\cL{{\cal L}}
\def\cS{{\cal S}}
\def\calD{{\cal D}}
\def\calV{{\cal V}}
\def\tilcalD{{\tilde{\cal D}}}
\def\bA{{\bf A}}
\def\bU{{\bf U}}
\def\bD{{\bf D}}
\def\tildebD{{\tilde{\bf D}}}
\def\bA{{\bf A}}
\def\tildebA{{\tilde{\bf A}}}
\def\bF{{\bf F}}
\def\tildebF{{\tilde{\bf F}}}
\def\bH{{\bf H}}
\def\G{{\rm G}}
\def\muG{{\mu{\rm G}}}
\def\nG{{\rm nG}}

\def\fun#1#2{\lower3.6pt\vbox{\baselineskip0pt\lineskip.9pt
  \ialign{$\mathsurround=0pt#1\hfil##\hfil$\crcr#2\crcr\sim\crcr}}}
\def\VEV#1{\left\langle #1\right\rangle}
\def\sigmav{\VEV{\sigma_{\rm ann}|v|}}
\def\vectheta{\mbox{\boldmath$\theta$}}
\def\vecl{\mbox{\boldmath $l$}}

\def\jcap{Jour. Cosmology and Astro-Particle Phys.}
\def\mnras{M.N.R.A.S}
\def\apj{Astrophys.J.}
\def\apjl{Astrophys.J.Lett.}
\def\apjs{Astrophys.J.Supp.}
\def\nat{Nature}
\def\na{New Ast.}
\def\nup{Nucl. Phys.}
\def\cmp{Comm. Math. Phys.}
\def\prl{Phys. Rev. Lett.}
\def\pl{Phys. Lett.}
\def\rmp{Rev. Mod. Phys.}
\def\ijmp{Int. Jour. Mod. Phys.}
\def\mpl{Mod. Phys. Lett.}
\def\pr{Phys. Rev.}
\def\prd{Phys.Rev.D}
\def\araa{Annu.Rev.Astron.Astrophys.}
\def\aap{Astron.Astrophys.}
\def\aaps{Astron.Astrophys.Supp.}
\def\aa{Astron.Astrophys.}
\def\pasj{Pub.Astr.Soc.Japan}
\def\pasp{Pub.Astr.Soc.Pacific}
\def\ssr{SpaceSci.Rev.}
\def\physrep{Phys.~Rep.}
\def\sovast{Soviet~Ast.}

\def\hl{\hline}
\def\tl{\tableline}

\begin{frontmatter}



\title{The Astrophysical Multimessenger Observatory Network (AMON)}


\author[psup,igc]{M.~W.~E.~Smith\corref{cauth}}
\ead{msmith@gravity.psu.edu}

\author[psua,igc]{D.~B.~Fox}
\ead{dfox@astro.psu.edu}

\author[psup,psua,igc]{D.~F.~Cowen}
\ead{cowen@phys.psu.edu}

\author[psup,psua,igc]{P.~M\'esz\'aros}
\ead{nnp@astro.psu.edu}

\author[psup,igc]{G.~Te\v{s}i\'{c}}
\ead{gut10@psu.edu}

\author[psua]{J.~Fixelle}
\ead{jxf5078@psu.edu}

\author[Columbia]{I.~Bartos}
\ead{ibartos@phys.columbia.edu}

\author[psup,psua,igc]{P.~Sommers}
\ead{sommers@phys.psu.edu}


\author[psup,igc]{Abhay~Ashtekar}
\ead{ashtekar@gravity.psu.edu}

\author[psus,psua]{G.~Jogesh~Babu}
\ead{babu@psu.edu}

\author[NASAGoddard]{S.~D.~Barthelmy}
\ead{scott.d.barthelmy@nasa.gov}

\author[psup,igc]{S.~Coutu}
\ead{coutu@phys.psu.edu}

\author[psup,igc]{T.~DeYoung}
\ead{deyoung@phys.psu.edu}

\author[psua,igc]{A.~D.~Falcone}
\ead{afalcone@astro.psu.edu}

\author[psup,psua,igc]{L.~S.~Finn}
\ead{LSFinn@psu.edu}

\author[psup,igc]{Shan~Gao}
\ead{sxg324@psu.edu}

\author[psup]{B.~Hashemi}
\ead{bth5032@psu.edu}

\author[Bonn]{A.~Homeier}
\ead{homeier@physik.uni-bonn.de}

\author[Columbia]{S.~M\'arka}
\ead{sm2375@columbia.edu}

\author[psup,igc]{B.~J.~Owen}
\ead{owen@gravity.psu.edu}

\author[GTech]{I.~Taboada}
\ead{itaboada@gatech.edu}


\cortext[cauth]{Corresponding author}

\address[psup]{Department of Physics, Pennsylvania State University,
  University Park, PA 16802, USA} 

\address[igc]{Institute for Gravitation and the Cosmos, Pennsylvania
  State University, University Park, PA 16802, USA} 

\address[psua]{Department of Astronomy \& Astrophysics, Pennsylvania
  State University, University Park, PA 16802, USA} 

\address[Columbia]{Department of Physics \& Columbia Astrophysics
  Laboratory, Columbia University, New York, NY 10027, USA} 

\address[psus]{Department of Statistics, Pennsylvania State
  University, University Park PA, 16802, USA}

\address[NASAGoddard]{NASA Goddard Space Flight Center, Code 661, Greenbelt, MD
  20771, USA}

\address[Bonn]{Physikalisches Institut, Universit\"at Bonn, Nussallee
  12, D-53115 Bonn, Germany}

\address[GTech]{Center for Relativistic Astrophysics, School of
  Physics, Georgia Institute of Technology, Atlanta, GA 30332, USA}


\begin{abstract}

We summarize the science opportunity, design elements, current and
projected partner observatories, and anticipated science returns of
the Astrophysical Multimessenger Observatory Network (AMON). AMON will
link multiple current and future high-energy, multimessenger, and
follow-up observatories together into a single network, enabling near
real-time coincidence searches for multimessenger astrophysical
transients and their electromagnetic counterparts. Candidate and
high-confidence multimessenger transient events will be identified,
characterized, and distributed as AMON alerts within the network and
to interested external observers, leading to follow-up observations
across the electromagnetic spectrum. In this way, AMON aims to evoke
the discovery of multimessenger transients from within observatory
subthreshold data streams and facilitate the exploitation of these
transients for purposes of astronomy and fundamental physics. As a
central hub of global multimessenger science, AMON will also enable
cross-collaboration analyses of archival datasets in search of rare or
exotic astrophysical phenomena.

\end{abstract}



\end{frontmatter}


\section{Introduction}
\label{sec.intro}

We stand at the dawn of multimessenger astrophysics -- a quest to use
the messenger particles of all four of nature's fundamental forces to
explore the most violent phenomena in the universe.  Observatories
first imagined a generation ago are finally being realized, including
the Advanced LIGO \cite{2010CQGra..27h4006H} and Virgo
\cite{2006CQGra..23S.635A} gravitational-wave detectors, the ANTARES
\cite{2011NIMPA.656...11A} and IceCube \cite{2010RScI...81h1101H}
high-energy neutrino observatories, and the Pierre Auger Cosmic Ray
Observatory \cite{2010NIMPA.613...29A}. On the ground and in space,
they are complemented by high-energy observatories including the
\swift\ \cite{2004ApJ...611.1005G} and
\fermi\ \cite{2012ApJS..203....4A} satellites, the HESS
\cite{2006A&A...457..899A}, VERITAS \cite{2011arXiv1111.1225H}, and
MAGIC \cite{2009arXiv0907.1211C} TeV gamma-ray telescopes, and the
HAWC \cite{2012JPhCS.375e2026T} TeV gamma-ray observatory.

Collectively, these facilities promise the first detections of
gravitational waves and high-energy cosmogenic neutrinos, the
resolution of the mystery surrounding the origins of ultrahigh-energy
cosmic rays, and a new window into the formation and evolution of
black holes. Given their nature as first-generation facilities,
however, the sensitivities of the non-electromagnetic observatories
are naturally limited, with rates of detection for transient events of
publishable significance known or expected to be low, perhaps a
handful per year (or, in the case of Advanced LIGO and Virgo at design
sensitivity, a few dozen \cite{2010CQGra..27h4006H}).

During the intervals prior to and between detection of these rare
high-significance events, the multimessenger facilities will be
buffeted by signals from a far greater number of lower-significance
events that will be statistically indistinguishable from background or
noise processes. Such ``subthreshold'' events are, by definition,
unrecoverable as astrophysical signals in the data stream of any
individual facility.  However, if they are accompanied by a
subthreshold signal in another multimessenger channel they can be
identified, and potentially achieve high significance, via careful
coincidence analysis of the data streams from multiple facilities.


In this paper we present the scientific case for the Astrophysical
Multimessenger Observatory Network and describe its important
elements.  AMON will weave together existing and forthcoming
high-energy astrophysical observatories into a single virtual system,
capable of sifting through the various data streams in near real-time,
identifying candidate and high-significance multimessenger transient
events, and providing alerts to interested observers.

As we show, AMON will enable a significant enhancement in the
effective aggregate sensitivity of the world's leading multimessenger
facilities for a small fraction of the facilities' total cost, provide
the first near real-time alerts for multimessenger transient sources,
and simplify the mechanics and politics of cross-collaboration
analyses for all partners. As such, we believe AMON represents a
natural next step in the extension of the global astronomical
community's vision beyond the electromagnetic (EM) spectrum.


The development of AMON is currently underway. Signatories to the AMON
Memorandum of Understanding\footnote{The AMON MOU is available at
  \url{http://amon.gravity.psu.edu/mou.shtml}} (MOU) include the
IceCube and ANTARES neutrino observatories, the HAWC TeV
\gray\ observatory, and the \swift\ orbital telescope. 
Exploratory discussions with the Pierre Auger Cosmic Ray Observatory, 
VERITAS, LIGO  (including
the GEO-600 collaboration), and \fermi\  scientific collaborations 
have led to signed letters of commitment,
with negotiations toward MOU signatures from all parties ongoing.
Discussions have also been initiated with candidate follow-up
facilities including \mbox{ROTSE-III} \cite{2003PASP..115..132A} and
the Palomar Transient Factory \cite{2009PASP..121.1395L}, with the
goal of bringing these observatories into the collaboration prior to
or shortly after the commencement of real-time AMON alert operations.

AMON is structured as an open and extensible network, with an MOU that
allows straightforward incorporation of new triggering and follow-up
facilities. Collaborations interested in the scientific goals of AMON,
and with useful triggering or follow-up capabilities to contribute,
are encouraged to contact the authors for information about joining
AMON.
First versions of the necessary supporting hardware and software
infrastructure for AMON are being installed at Penn State, and initial
analyses on archival and simulated real-time data streams will get
under way shortly thereafter, as a means of preparing to bring the
first set of triggering facilities on-line within the next year.


The paper is structured as follows: In Section~\ref{sec.opp} we
provide an overview of the scientific opportunity for AMON which
motivates our efforts. In Section~\ref{sec.ops} we discuss the
elements of AMON, including the technical and operational protocols
that we propose to adopt in linking the partner facilities, and the
algorithms that we will use to identify coincident
signals. Section~\ref{sec.power} presents detailed simulations of
multimessenger transient sources, and additional theoretical case
studies, which illustrate the gains that stand to be realized by
AMON. Section~\ref{sec.conclude} presents our summary and conclusions.

\section{The AMON Science Opportunity}
\label{sec.opp}

AMON is intended to contribute in several ways to the first decade of
multimessenger astronomy. A common focus of these approaches is on
multimessenger transient events that are observed as coincident
(potentially subthreshold) signals in the data of AMON partner
facilities corresponding to two or more distinct types of messenger
particle. In this section, we briefly review the strongest candidates
for these multimessenger transient sources and the current theoretical
expectations for their properties, rates, and broader implications for
physics and astrophysics.


\subsection{High-Luminosity Gamma-Ray Bursts}
\label{sub.opp.hlgrbs}

As the most violent cataclysms known, and as the sites of the most
highly-relativistic cosmic outflows, gamma-ray bursts (GRBs) have long
been considered likely multimessenger transient sources. Recent
progress in distinguishing the distances and energetics of GRBs and
understanding the likely nature of their
progenitors~\cite{2009ARA&A..47..567G} means that we can review the
expected multimessenger signals of the different varieties of GRB on a
case by case basis.


The traditional or ``classical'' long-duration, high-luminosity
gamma-ray bursts (HL-GRBs) are the most common type of GRB detected by
satellite experiments, being observed as bright seconds- to
minutes-long bursts of $\gamma$-radiation from high-redshift, $z\simgt
1$. HL-GRBs are believed to arise when a massive star ($M \simg
25\msun$) undergoes core collapse to a black hole (BH); confirmation
of this ``collapsar'' model~\cite{1999ApJ...524..262M} for the HL-GRBs
has been most dramatically provided by spectroscopic observations of
ensuing type Ibc supernovae (SNe), with other lines of evidence also
contributing~\cite{2006ARA&A..44..507W}.

In the collapsar model, formation of a high angular-momentum BH and
accretion of residual gas through an accretion disk produces a
relativistic jet. In ``successful'' GRBs the jet pierces the stellar
envelope, accelerates to high Lorentz factor, and radiates gamma-rays
for tens of seconds, providing a bright electromagnetic trigger for
observers within the jet collimation angle. The typical HL-GRB jet
energy of $E\sim 10^{51}$ to $10^{52} \erg$, collimated within an
angle of $\VEV{\theta}\sim 5\arcdeg$, roughly 1/250 of the sky, yields
the observed isotropic-equivalent energies of $E_0=E\,(4\pi/\Omega_j)
\sim 10^{53}$ to $10^{54}\erg$ \cite{2011ApJ...732...29C}.  For a
burst duration $T\sim 10\s$, this corresponds to an
isotropic-equivalent luminosity $L_\gamma \sim E_0/T =10^{52}$ to
$10^{53}\erg\s^{-1}$. Given a 250:1 jet collimation factor, the
nearest off-axis bursts are anticipated to be located $250^{1/3}\sim
6$ times closer than the nearest on-axis bursts. Off-axis bursts,
defined as bursts whose bright initial \gray\ emission does not
illuminate observers at Earth, may still be observed via their less
luminous shock-breakout emission, their prompt gravitational wave (GW)
signal, or their subsequent afterglow and/or supernova components.

In the standard internal-shock model, the gamma-ray emission of
HL-GRBs is produced by internal shocks at a dissipation radius
$r_{d}\sim \Gamma^2 c\,\delta t \sim (3\times 10^{13} \, {\rm cm}) \,
\eta_{2.5}^2 \, \delta t_{-2}$~\cite{2006RPPh...69.2259M}, where
$\eta_{2.5}=\Gamma/300$ is the bulk Lorentz factor divided by 300 and
$\delta t_{-2}=\delta t/10^{-2}\s$ is the variability time of the
central engine in hundredths of a second.  These shocks
Fermi-accelerate electrons, which produce gamma-rays via synchrotron
and inverse Compton scattering, and are then boosted by the bulk
motion of the relativistic outflow.  For a discussion of alternative
models see ref.~\cite{2012arXiv1204.1897M}.


The same shocks responsible for electron acceleration and $\gamma$-ray
emission should also accelerate protons, leading to photo-produced
pions which in turn yield high-energy cosmogenic neutrinos (\HEN) and
$\gamma$-rays from charged and neutral pion decays, respectively
\cite{1997PhRvL..78.2292W,1998PhRvD..58l3005R}.  Initial assumptions
were that the energy in relativistic protons would be of the same
order as the energy emitted in $\gamma$-rays, $E_p \simeq (1/f_e)
E_\gamma$, where $f_e \siml 1$ is the fraction of proton energy given
to electrons (and observed as GRB photons), so that the optical depth
$\tau_{p\gamma}$ determines the \HEN~and TeV gamma-ray luminosity,
$L_{\rm TeV} \sim \tau_{p\gamma}L_p$.  The \HEN~to $\gamma$-ray flux
ratio expected in this model has been quantified
\cite{2004APh....20..429G} and used by IceCube
\cite{2011PhRvL.106n1101A, 2011PhRvD..84h2001A, 2011APh....35...87A}
to set limits which are already a factor of five below the naive
predictions. However, careful consideration of the underlying physics
\cite{2011arXiv1112.1076H, 2012PhRvD..85b7301L, 2012arXiv1204.0857H}
suggests that IceCube observations will need to continue for several
more years before the somewhat reduced \HEN~fluxes of more realistic
models are tested. Ultimately, detection of GRB-related \HEN~is
expected if HL-GRBs contribute a significant fraction of the
highest-energy cosmic rays.


Regardless of the total energy release of the HL-GRBs, which may well
be an order of magnitude or more above the beaming-corrected
$\gamma$-ray energies $E\simlt 10^{52}\erg$, no significant GW
emission will be produced if the core collapse, jet production, and
burst processes maintain approximate axisymmetry throughout. However,
it is possible for a rapidly rotating core and accretion disk to
develop bar and/or fragmentation instabilities which could result in
substantial GW emission. In the most optimistic
case~\cite{2003ApJ...589..861K} the resulting GW signal is periodic
and roughly as strong as the signal from neutron star binary mergers,
and hence, visible out to similar distances (hundreds of Mpc for the
advanced facilities) using ground-based GW detectors.

While some 3D simulations of stellar collapse suggest a fragmentary
process, and a correspondingly weaker GW energy release $E_{GW}\sim
10^{-7}\msun c^2 \sim 10^{47}\erg$~\cite{2011PhRvL.106p1103O}, these
have not included relativistic effects that are known to contribute to
large bar asymmetries~\cite{2009ApJ...702.1171C}, as manifested in
other simulations~\cite{2011PhRvL.106y1102K}. When present, these
instabilities result in GW emission comparable to that found in
optimistic analytical calculations~\cite{2003ApJ...589..861K}.


\subsection{Low-Luminosity Gamma-Ray Bursts}
\label{sub.opp.llgrbs}

Low-luminosity GRBs (LL-GRBs) are underluminous long-duration GRBs,
having substantially lower isotropic-equivalent energies, $E_0\sim
10^{49}$ to $10^{50}\erg$, than the HL-GRBs.  Because of their lower
$\gamma$-luminosities, they are currently only detected at low
redshift, $z\siml 0.5$, where they provide the bulk of the observed
GRB-SN Ibc associations~\cite{2006ARA&A..44..507W}. Estimates of the
LL-GRB rate suggest that they occur at $\simgt$100 times the rate of
HL-GRBs while constituting $\simlt$1\% the rate of type Ibc SNe
overall
\cite{2006Natur.442.1014S,2007ApJ...657L..73G,2007ApJ...662.1111L,2010Natur.463..513S}.

The $\gamma$-ray emission of LL-GRBs may be due, in some or all cases,
to a relativistic shock breakout
\cite{2010ApJ...716..781K,2012ApJ...747...88N} and hence may not
require the Lorentz factor $\Gamma\simgt 100$ jet needed to explain
the high-luminosity, high-variability $\gamma$-emission of
HL-GRBs. Such shock breakout emission would also be expected to be
uncollimated (uniform over the sky) or nearly so.


Predictions for the \HEN\ emissions of \mbox{LL-GRBs} have been
explored by \cite{2006ApJ...651L...5M,Gupta+07grbnu} for relativistic
jet models and by \cite{2012arXiv1210.8147K} for shock breakout
models. In relativistic jet models, a straightforward approach scales
\HEN\ fluxes from HL-GRB predictions according to their $\gamma$-ray
luminosities, and anticipates similar spectra with $E_\nu^2 \phi_\nu$
peaking at PeV to EeV energies. Given current evidence for
less-relativistic jets in LL-GRBs, however, these predictions may be
optimistic.

Shock breakout models \cite{2012arXiv1210.8147K}, by contrast, predict
softer neutrino spectra peaking in the TeV to PeV range, with
luminosities for reference events that would make individual bursts
detectable to IceCube within a horizon of $D\simlt 10$\,Mpc. The
\mbox{LL-GRB} rate within this horizon is thought to be small, $r\sim
0.002$\,\peryear, but prospects for detection could be improved by
stacking analyses using $\sim$dozens of more distant events identified
by their prompt high-energy EM signature.


Similarities in explosion energy, ejecta velocity, and synthesized
nickel mass for the type Ibc supernovae of LL-GRB and HL-GRB events
suggest that the details of core collapse for the two event classes,
and their GW emissions, may be similar. However, given the critical
role that high values of core angular momentum are thought to play in
powering HL-GRB jets, and the need for accretion or disk instabilities
to power significant GW emission, the relative weakness of LL-GRB jets
may suggest correspondingly reduced prospects for GW emission. If the
GW emissions are competitive with those of binary neutron star-neutron
star (NS-NS) mergers in even some cases, this would make the LL-GRBs
with their relatively isotropic \gray\ emissions a highly-promising
target population for ground-based GW detectors.


\subsection{Short-Hard Gamma-Ray Bursts}
\label{sub.opp.shbs}

Short-hard gamma-ray bursts (SHBs), apart from their shorter durations
and somewhat harder spectra, are observationally similar to
HL-GRBs. Their harder spectra may indicate higher bulk Lorentz factors
\cite{2006RPPh...69.2259M}, while their reduced durations, $t_b\siml
2\s$, suggest shorter accretion timescales compared to those of the
HL-GRB collapsars. A softer ``extended emission'' episode lasting for
$t\sim 30$ to $100\s$ after the burst itself is present in about a
third of SHBs, accounting for 5-20\% of the total energy
\cite{2006ApJ...643..266N}.

Thanks to afterglow discoveries and host galaxy identifications of the
\swift\ era, consensus now holds that SHBs are likely due to compact
binary (NS-NS or NS-BH) mergers
\cite{2007PhR...442..166N,2009ARA&A..47..567G}. If so, these events
are associated with progenitor systems that are observed within our
own Galaxy (as relativistic NS-NS pulsar binaries), have merger rates
that can be estimated from population synthesis modeling (e.g.,
\cite{2004ApJ...601L.179K}), and should emit strong GW emission in a
highly-calculable ``chirp'' waveform across the frequency range of
ground-based GW observatories just prior to merger. Indeed, binary
mergers are the primary extragalactic target population for
next-generation LIGO and VIRGO, with expected event rates at full
design sensitivity expected to reach dozens annually
\cite{2010CQGra..27h4006H}.

Afterglow observations for a handful of \swift-detected SHBs show
evidence for typical collimation angles $\VEV{\theta}\simgt 6\arcdeg$,
corresponding to beaming fractions of 200:1 (similar to that for
HL-GRBs) or less~\cite{2012ApJ...756..189F}. These beaming corrections
adjust the observed isotropic-equivalent energies of the bursts,
$E_0\simgt 10^{51}\erg$ to lower inferred jet energies $E \sim
10^{48}$ to $10^{50}\erg$. As such the nearest off-axis mergers are
predicted to be observed at $200^{1/3}\sim 6$ times closer distance
than the nearest on-axis events, where they may be detected as GW
inspiral signals without SHB counterparts.


Because of their very similar $\gamma$-ray characteristics and
inferred jet properties, it is assumed that the radiation physics of
the HL-GRBs and SHBs are similar, apart from possible slight
differences due to the SHBs' reduced durations and potentially larger
Lorentz factors and collimation angles
\cite{2006RPPh...69.2259M}. \HEN~production models may thus
conservatively be carried forward by assuming similar \HEN~to
$\gamma$-ray flux ratios as for the HL-GRBs.


In terms of GW emission, SHBs in the compact binary merger model
represent a ``dream scenario'', as systems in which the GW waveform is
known to be strong (detectable to hundreds of Mpc by advanced
detectors) and calculable, and moreover, has already been implemented
in data analysis systems via matched-filter algorithms that will be
run in real-time by LIGO and VIRGO, enabling the observatories to
distribute GW ``inspiral alerts'' to interested observers. The primary
questions for the near future, then, are whether the GW inspiral
signatures of these mergers will be detected, at what rate, and whether
or not a coincident $\gamma$-ray or other electromagnetic signature
can be identified. Unless our understanding is rather radically
misplaced, all three questions are likely to be resolved once the era
of the advanced GW facilities is fully under way.


\subsection{Choked Jet Supernovae}
\label{sub.opp.choked}

The choked jet supernova, a theoretical scenario, represents the
alternative fate that awaits a massive star if its core collapse
generates a high-energy jet as in HL-GRBs, but in a fashion or within
the context of a higher-mass stellar envelope that absorbs the jet
energy before it is able to escape the star. As such, the choked-jet
events evince no high-luminosity $\gamma$-ray emission; however, in
the course of being quenched, their sub-stellar jets may undergo
internal shocks which could accelerate protons and yield $p\gamma$ and
$pp$ neutrinos in the TeV range
\cite{2001PhRvL..87q1102M,2003PhRvD..68h3001R,2008PhRvD..77f3007H}.
Observationally, then, the choked-jet supernovae would be observed as
a relatively low-redshift, highly-energetic supernova (a hypernova)
with associated \HEN~emission.

Event rate estimates for choked-jet events are necessarily
speculative. However, we note that if the process that leads to
high-energy jet production during core collapse is largely independent
of the nature of the overlying stellar envelope, then the choked-jet
supernova rate may be comparable to the rate of HL-GRBs.  Moreover,
given that the collapsar mechanism has multiple failure modes,
including a high-mass stellar envelope, insufficient rotation, and too
much jet precession~\cite{2004ApJ...608..365Z}, one can speculate that
choked jet supernovae should, on generic grounds, be more common than
HL-GRBs, although potentially (for these and related reasons)
exhibiting less-luminous \HEN~and/or GW emission.


\subsection{Core Collapse Supernovae} 
\label{sub.opp.ccsne}

Core-collapse supernovae (CCSNe) are expected roughly every 30 to 100
years in our Galaxy; given the direct heritage of the revolutionary
neutrino detection of SN\,1987A~\cite{1989ARA&A..27..629A}, any such
Galactic cataclysm would obviously represent a first-grade candidate
multimessenger transient.

Seen in relatively nearby galaxies (distances $D\simlt 100$\,Mpc),
CCSNe are typically detected by optical observers within a few days,
although as \swift\ has demonstrated~\cite{2008Natur.453..469S},
prompt detection of the high-energy shock breakout can be realized
with sufficiently sensitive instrumentation. On this timescale, the
prompt thermal ($E_\nu\simgt 10$\,MeV) neutrinos should be detectable
by IceCube and Super-Kamiokande for events within $D\simlt 50$\,kpc
\cite{2011A&A...535A.109A}.

\HEN\ may be produced by these ``ordinary'' CCSNe, either at shock
breakout \cite{2012arXiv1210.8147K} or at later times, via interaction
of the SN shockwave with a dense circumstellar medium
\cite{2012arXiv1206.0748O}. The shock breakout \HEN\ emissions of
ordinary CCSNe will be fainter than for the LL-GRBs, and so require
very nearby events or larger-scale stacking analyses to detect with
IceCube. CCSNe circumstellar interactions have the potential to
generate a greater number of \HEN\ over an extended timescale of weeks
to months, and may be detectable with IceCube, if candidate
\HEN\ CCSNe can be identified with confidence and in sufficient
numbers to make a sensitive search.

GW emission from generic CCSNe are expected to be relatively weak
\cite{2011PhRvD..83f4008R}, and hence, not detectable except in the
case of a Galactic (or possibly, Small or Large Magellanic Cloud)
event.

Cosmic rays with energies below the ``ankle'' of the spectrum at
$E\approx 4\times 10^{18}$\,eV are generally attributed to sources
within the Galaxy~\cite{2006astro.ph..7109H}.  Since no steady sources
have been detected despite sensitive searches
\cite{2011arXiv1107.4805T}, they may be produced in occasional bursts
within the Galaxy, including Galactic CCSNe. The decay length for
relativistic neutrons is $9.2\,E_{18}$\,kpc, where $E_{18}$ is the
neutron energy in EeV.  Transient sources of neutrons of 1~EeV or
above can thus be detected across much of the Galactic disk, including
the central bulge region, and at higher energies also accessible with
Auger, sources throughout most of the halo are detectable.


\subsection{Blazars}
\label{sub.opp.blazars}

Blazars are frequently detected in $\gamma$-rays at GeV (Fermi, AGILE)
and TeV (HESS, VERITAS, MAGIC) energies; they are also highly luminous
\xray, optical, and radio sources. Since the non-thermal
\xray\ emission of blazars is generally understood to be synchrotron
emission from electrons accelerated in shocks of an inner jet,
shock-accelerated protons are expected as well. The target photons for
$p\gamma$ interactions, leading to \HEN~emission, could then be either
synchrotron photons from co-accelerated electrons, or optical/UV
photons originating in the accretion disk or scattered into the jet by
broad-line region clouds.

Blazars are highly variable, flaring sources that are much brighter
across the EM spectrum during flare episodes than in
quiescence. Expected \HEN\ fluxes for standard blazar models
\cite{1998PhRvD..58l3005R,2001PhRvL..87v1102A,2003APh....18..593M}
suggest that typical individual flares cannot be detected with
IceCube, with the summed contribution of $\simgt$100 such flares
required to yield an expected $>$1 \HEN\ detection in
IceCube. Separately, limits on the diffuse \HEN\ flux due to the
summed contribution of all blazars in the Northern hemisphere have
been published by IceCube using the 40-string (roughly one year
integration) dataset, constraining some models
\cite{2011PhRvD..84h2001A}, but without the statistical leverage that
would be provided by comprehensive EM monitoring yielding the times
and durations of blazar flares.

Individual spectral components of blazar flares are also variable, and
the properties of the highest-energy components are poorly constrained
due to the difficulty in obtaining sensitive TeV \gray\ data and
simultaneous flare-triggered multiwavelength EM coverage. Moreover,
recent \fermi\ and multiwavelength data suggest that spectral breaks
and possible additional components are present during some blazar
emission episodes
\cite{2010ApJ...716...30A,2010ApJ...717L.118P,2010arXiv1006.5048B}. 
Exceptional TeV \gray\ flares from blazars that lack associated
\xray\ emission, analogs of the 1ES~1959+650 ``orphan'' TeV flare
\cite{2003ApJ...583L...9H,2004ApJ...601..151K,2005APh....23..537H},
could be associated with hadron acceleration that would yield
\HEN\ fluxes well in excess of those expected from typical flares,
and hence, would be more readily detected by current facilities.

Energetic GW emission from blazars in the frequency range of
ground-based detectors is not anticipated.


\subsection{Primordial black holes and other exotica}
\label{sub.opp.exotic}

If primordial black holes (PBHs) formed in the early universe with
masses $M\sim 5\times 10^{14}$\,g appropriate for them to undergo
explosive evaporation at the present epoch \cite{1974Natur.248...30H},
then they would serve as a distinct and exotic type of multimessenger
transient.

As the black hole loses mass via Hawking radiation, $dM/dt =
-\alpha(M)/G^{2}M^{2}$~\cite{1976PhRvD..13..198P}, its temperature
$T_{\rm BH}$ increases, allowing an increasing number of particle
types (degrees of freedom or {\it dof\/}) $\alpha(M)$ to be radiated.
Once $T_{\rm BH}$ reaches the quantum chromodynamic energy scale of
$\Lambda_{\rm {QCD}}\simgt 200$\,MeV, free quarks and gluons will be
emitted and fragment into hadrons, photons and leptons, resulting in a
flux of high-energy cosmic ray neutrons and
\HEN~\cite{1990PhRvD..41.3052M} that might be detectable from Auger
and IceCube.

Detection of PBHs would give dramatic confirmation of Hawking's
hypothesis and theories positing early cosmological phase transitions
(e.g.,~\cite{1967SvA....10..602Z,1971MNRAS.152...75H}), and enable
studies that would likely yield deep insights into ultrahigh-energy
physics as well as quantum gravity.


Other primordial relics, if they exist, might have decay modes
yielding harder spectra than the PBHs
(e.g., \cite{1983NuPhB.224..469H,1996APh.....4..253P,1997PhRvL..79.4302B,1998PhRvD..58j3515B,2002NuPhB.621..495S,2006PhRvD..74b3516A,2006PhRvD..74k5003E}).
Power-law spectra with $n\propto E^{-2}$ or harder would put out a
majority of the total decay energy at the highest energies, meaning
that the first detections, or most sensitive limits, on these
processes are likely to arise from the highest-energy (largest area)
facilities, IceCube and Pierre Auger.


\section{AMON Dataflow and Operations}
\label{sec.ops}

Multimessenger detection of one of the astrophysical sources described
above, or of some entirely distinct cosmic phenomenon, will require
coordination and cooperation between observatories of strikingly
different design and operation.  These observatories produce data
whose heterogeneity motivates the unified approach of AMON for
detecting coincidences.  Individual facilities participating in AMON
can be characterized as triggering facilities, follow-up facilities,
or both.  Triggering facilities are typically sensitive to one or more
messenger type (photons, cosmic rays, neutrinos, or gravitational
waves) and search in a wide field of view or monitor known sources for
transient behavior. From their raw data, they construct candidate
astrophysical events, here denoted ``trigger events'', which may
transmitted through either public or private channels to AMON. These
variously represent a single particle interaction ({\it e.g.}, a muon
neutrino interaction detected by IceCube), multiple detections
processed into a high level trigger ({\it e.g.}, a Swift BAT light
curve), or a sharp rise in a continuous measurement ({\it e.g.}, an
Advanced LIGO event).

For AMON, a fundamental characterization of each observatory is its
event-wise false positive rate (FPR).  The FPR is the dominant
component of the total event rate that each observatory sends to AMON.
The FPR is typically due to intrinsic detector noise or irreducible
backgrounds that produce signal-like events.  The FPRs for all
triggering observatories considered here are listed in
Table~\ref{table.FPR}a.  Later in this section we describe
how these FPRs are calculated.

By design, AMON will handle data streams dominated by these
subthreshold signal-like events that individually cannot rise to the
level of astrophysical discovery. For example, a single muon neutrino
detected by IceCube could be of astrophysical origin, but the channel
is indistinguishable from, and dominated by, atmospheric muon
neutrinos.  Similarly, single Galactic neutrons in Auger cannot be
distinguished from the dominant background of charged particle cosmic
ray events, and lowering the threshold of \xray, \gray, or
gravitational wave detectors increases signal acceptance at the cost
of simultaneously admitting many additional accidental noise events.
However, if two or more of these events are correlated in arrival
direction and time they can attain a combined significance that
results in AMON issuing an ``Alert.''  An Alert either constitutes an
immediate discovery or prompts follow-up observations that could
demonstrate the presence of a ``smoking gun'' electromagnetic
counterpart.  By having all individual observatory FPRs on hand,
algorithms running in the AMON framework are uniquely capable of
calculating the aggregate Alert FPR associated with each coincidence
detection.  The Alert FPR is a statistical measure of the quality of a
given multimessenger coincidence, endowing each Alert with a high
degree of utility for multimessenger source searches.

It is also possible for trigger events from a single observatory to be
intrinsically above threshold, allowing the observatory to make a
stand-alone claim for discovery.  A burst of three neutrinos in close
directional and temporal coincidence (IceCube or ANTARES), a burst of
three air showers in coincidence (Auger), especially if their putative
source is close enough for them to be neutrons, a statistically
significant burst of hard X-rays (Swift BAT, Fermi GBM) or
$\gamma$-rays (Fermi LAT or HAWC), or a strong strain registered by a
gravitational wave detector (LIGO-Virgo) would each constitute a
discovery.  For these stand-alone discoveries, AMON can serve as a
quick and convenient conduit for disseminating the source coordinates,
enabling timely follow-up observations at multiple disparate
observatories.  AMON will leverage the GRB Coordinates Network
(GCN)~\cite{GCN}, among others, to promulgate its Alerts.

\begin{figure}[h]
\centering
\includegraphics[width=0.49\textwidth]{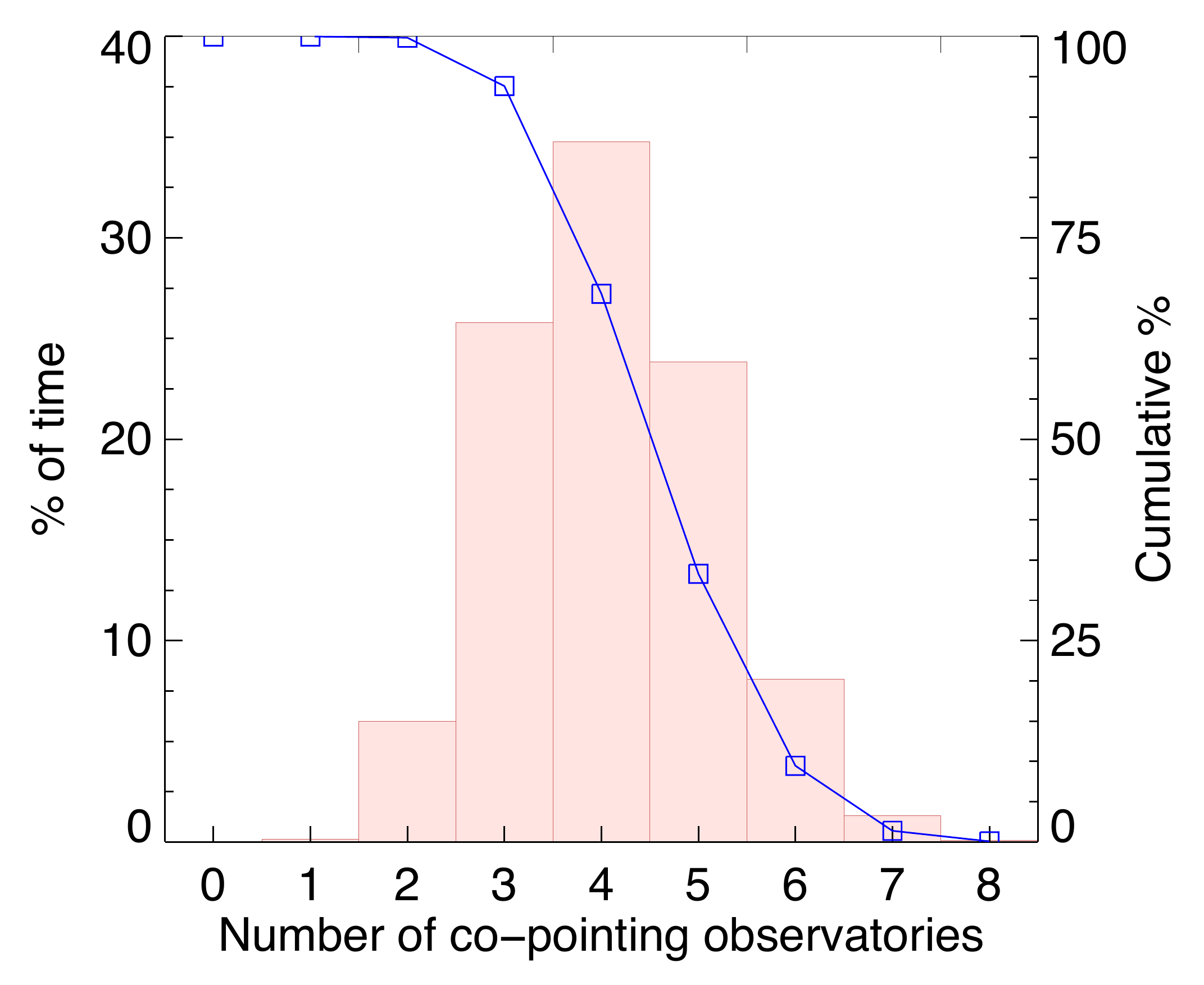} 
\includegraphics[width=0.49\textwidth]{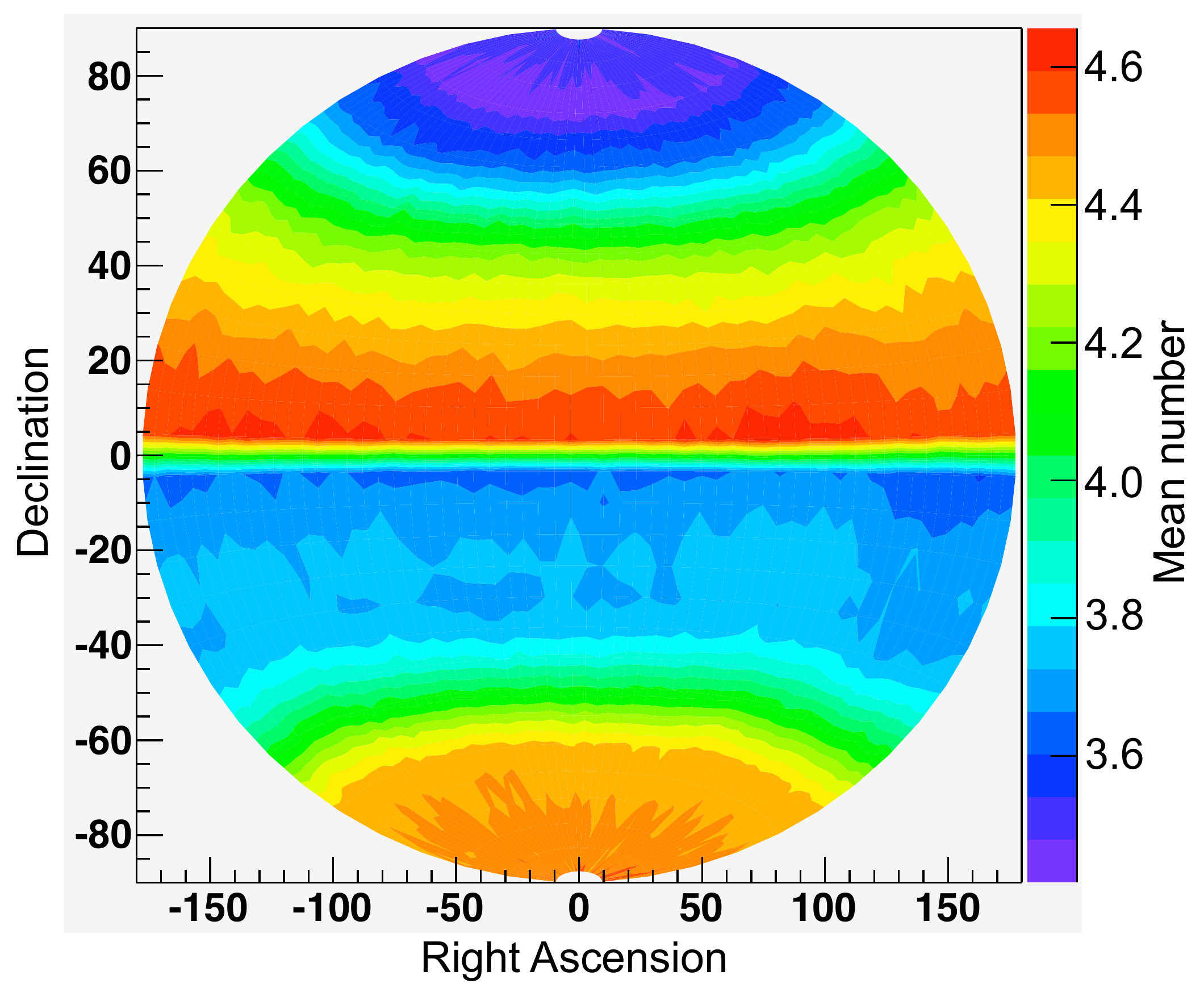}
\caption{\small (left) Distribution of the total number of trigger
  facilities observing a cosmological source, averaged over time and
  sky location.  (right) Average number of observatories
  simultaneously viewing a source, as a function of location.  Spatial
  variance is caused by the limited field-of-view of some
  observatories and the movement of orbital telescopes.  }
\label{fig.mult}
\end{figure}

Underpinning AMON's sensitivity to new phenomena is the broad
(multiple steradian) sky coverage and high duty cycles of the
triggering observatories.  Fig.~\ref{fig.mult} shows the results of a
simulation of a calendar year for the triggering facilities that have signed the 
AMON MOU (IceCube, ANTARES, HAWC, and Swift  BAT) and potential future
signatories (Auger, Fermi LAT and GBM, and LIGO-Virgo), establishing that -- absent
severe disruptions -- at least two facilities are always observing any
given part of the sky, and that about 94\% of $4\pi$~sr$\cdot$yr lies
within the field of view of three or more facilities.

Existing multimessenger searches involving only pairs of observatories
are typically in a master-slave relationship,  {\it e.g.}, a triggering
observatory initiates follow-up at an optical
telescope~\cite{2012A&A...539A..60A,2012arXiv1205.1124E,2011ICRC....4..185H}.
In contrast, AMON will enable archival and real-time searches for
coincidences among multiple observatories in a peer-to-peer
relationship that will provide markedly increased opportunity for
discovery of new multimessenger sources, as detailed in Sec.~4 below.
The real-time AMON analysis will be informed by the extensive effort
already invested by the triggering
observatories~\cite{2012A&A...539A..60A,0004-637X-701-2-1721,
  Braun2010175,2008PhRvD..77f2004A, 2010ApJ...715.1438A,
  2007PhRvD..76f2003A, abbott:211102, 2011ApJ...734L..35A, LIGO_EM,
  2005SSRv..120..143B, Abdo2010}.  The trigger events these
observatories produce will be heterogeneous, covering the full range
of messenger particles and differing significantly in background rate
and localization precision. In some cases the events will include
additional quality information that may be useful in refining the
single-stream false positive rate. For purposes of AMON, all events
will be couched in the common language of statistics, containing
information including a trigger time, event position, and a positional
error or probability density function (PDF).

 Once the individual event streams have been characterized, including
 tracking observatory pointing and FPR, the multiple event streams can 
 be combined.  At the most basic level, this will be carried out via a
 coarse clustering analysis, searching for temporal and spatial
 coincidences.  The output of the clustering analysis is dominated by
 pairwise coincidences, with the FPR for an ordered pair of triggers from
 observatories $(a,b)$ given by\footnote{In some cases, it
  may be advantageous to use one event stream to trigger event
  selection for another, {\it e.g.}~if the rate of single events is
  high, as with HAWC, or if the localization is poor, as with
  low-significance gravitational wave events. However, processing of
  an observatory's data can always be carried out, either at AMON or
  prior to transmission, to yield a manageable single-observatory FPR,
  and such preprocessing has been assumed for this discussion.},
\begin{equation}
\FPR{ab} \approx \frac{R_a}{\Omega_a} \frac{R_b}{\Omega_b}
\Delta T\langle\Omega_{ab}\rangle\Delta\Omega_{ab},
\label{eqn.RFP1}
\end{equation}
where $R_a$ is the subthreshold event rate from observatory $a$ to
AMON (assumed constant here), $\Omega_a$ is the field of view of
observatory $a$, $\Delta T$ is the total width of the temporal search
window, $\langle\Omega_{ab}\rangle$ is the time-averaged overlapping
field of view of observatories $a$ and $b$, and $\Delta\Omega_{ab}$ is
the search area for coincidences between observatories $a$ and $b$.
In Table~\ref{table.FPR} (row ``b''), we show the result of this
calculation for $\Delta T = 100$ s and $\Delta\Omega_{ab}$ adjusted to
give $\sim90$\% acceptance for signal event pairs (where each event is at
the threshold for transmission to AMON).  The self-coincidence of the 
individual observatories is given by the diagonal $R_{aa}$ (although this 
calculation is not appropriate in the case of preprocessed event streams) and the 
sum $R_{ab} + R_{ba}~(a \ne b)$ gives the total number of pairs from 
$a$ and $b$ without respect to order.   
As expected, the combined
FPR due to a pair of false positive events is orders of magnitude
lower than that of the individual component subthreshold event
streams.

This particular calculation is by no means definitive.  A trivial
adjustment of the cuts will increase the rates (useful for generating
AMON Alerts for testing purposes). More importantly, additional
information can drastically lower the FPR for the AMON analysis
(Table~\ref{table.FPR}, row ``c'').  For example, one can
statistically stack the Alerts though an archival study, or shorten the
temporal search window to search for exotic phenomena. In real-time,
one can require that at least one event be of high significance or
that the coincidence contains at least three events\footnote{The 
FPR for a coincidence of the ordered triple $(a, b, c)$ is given by 
  $\FPR{abc} \approx  \frac{1}{2}\frac{R_a}{\Omega_a}\frac{R_b}{\Omega_b}
  \frac{R_c}{\Omega_c}\left(\Delta T\Delta\Omega_{abc}\right)^2\langle\Omega_{abc}\rangle$,
  where the factors $\Delta\Omega_{abc}$, an equivalent search area for a threefold
  coincidence, and $\langle\Omega_{abc}\rangle$, the joint field of view, are determined by 
  Monte Carlo. The factor of $\frac{1}{2}$ depends on how the temporal search is
  defined, where here we require that all three events occur within $\Delta T$. In 
  Table~\ref{table.FPR}c, we present the totals $\sum_{bc}\FPR{abc}$, for the case
  where all Auger data is included and for the case where Auger data is restricted
  to the galactic plane.}.

A promising feature of AMON is its ability to distribute Alerts in
real-time to initiate follow-up observations.  The relatively high FPR
represented by the sum total of all pairwise coincidences among the
observatories considered (Table~\ref{table.FPR}, row ``b'') may be
tolerable for some follow-up facilities, especially if their response
is highly automated. However, many follow-up facilities will have
limited observing time available for this science and may require a
substantially lower FPR for triggering.  This can be achieved via a
refined analysis, providing a combined best fit location and error, as
well as a maximum likelihood ratio or Bayesian probability. The
likelihood or probability measure is drawn from a continuous
distribution, in contrast to the FPR of the clustering analysis which,
given a fixed event rate for each stream and a fixed set of cuts,
takes on a unique value for any pair of event streams.  Prior
information can easily be incorporated into the likelihood analysis
including, for example, previous detection limits, observatory
sensitivity, or a galaxy catalog, and these choices can be tuned for
different follow-up programs.

As an independent channel of scientifically-relevant data and
constraints, follow-up observations have the potential to provide the
added information required for a definitive discovery.  By
distributing candidate source positions in real-time to follow-up
facilities, with statistically valid measures of the FPRs, AMON will
enable efficient fast-response counterpart searches and studies across
the electromagnetic spectrum. The existence and nature of such EM
counterparts may prove decisive in verifying the existence of some of
the first multimessenger sources, and can reveal the nature and
detailed properties of the source, including important ancillary
information such as the source environment, host galaxy (if any), and
properties of intervening gas and dust along the line of sight.  The
discovery potential of pairwise coincident events with positive
follow-up detection is described in more detail below.

AMON will be deployed in three phases. During the first year of
operation beginning in mid-2013, AMON will use archived data to
develop and tune analysis algorithms. Such archival analyses, while a
necessary first step prior to activation of real-time alerts, will
also be of intrinsic scientific interest, and may point to possible
new astrophysical signals or new and constraining upper limits on
jointly-emitting source populations.  Even as AMON moves toward
real-time operation, certain source populations will be best
discovered through a collaborative blind study of accumulated data;
for example, in searches for exotic particles, as discussed below for
primordial black holes.  To enable this, participating AMON members
will, at any time, be able to utilize the AMON database and data
products, under rules that are established in the AMON MOU~\cite{MOU}.

In subsequent operational phases, standard AMON analyses will be
modified to operate in real-time, enabling identification of candidate
transient sources as soon as the events are transmitted. At first, the effort will focus on
those discoveries that can be made by the triggering facilities alone,
with FPRs similar to those in Table~\ref{table.FPR}c. Even at this
stage, the real-time nature of the network plays a critical role,
providing the triggering facilities with AMON Alerts that can prompt a
closer analysis of their own data. The Alerts shared at this stage
will include many whose combined FPR is too high to claim detection of
a transient.  These will not only exercise the AMON system, but may
initiate a deeper search by those participating observatories that did
not initially report a triggering event. An example of this discovery
mode is the joint search for high-energy cosmogenic neutrinos and
gravitational waves, as discussed below.
    
As quickly as is feasible, AMON will transition to its final phase,
with Alerts distributed to participating follow-up observatories to
enable the near real-time search for electromagnetic counterparts. For
sources that exhibit detectable afterglow, this is the most powerful
technique, leveraging the high acceptance of the initial FPR
threshold, yet resulting in a final FPR that is lower than that
available by the other techniques. The estimated final FPRs are 
shown in Table~\ref{table.FPR}c, for the case of serendipitous 
follow-up detection of GRB afterglow (assuming Swift XRT sensitivity
and a $\sim1^\circ$ search region) or a SNe light curve (assuming the 
detection limits of \cite{2012A&A...539A..60A}). 
In the next section, as a proof of concept, we show that significant sensitivity
gains are made by AMON in the search for electromagnetic counterparts
to cosmogenic neutrino sources.

\begin{landscape}

\def\na{{N/A}}

\ctable[
caption={\small (a) False Positive Rate (FPR) in units of yr$^{-1}$ for 
single event streams, including {\it above threshold} events that can lead to 
stand-alone discoveries and the background-dominated {\it subthreshold} 
events that are transmitted to AMON. (b) A coarse clustering analysis, shown 
here for a $3\sigma$ spatial window and  $\Delta T = 100$ s, will primarily 
identify pairs of coincident subthreshold events with corresponding pairwise 
FPR. This FPR can be further reduced with a likelihood analysis to tune the 
distribution rate of AMON Alerts, with the subsequent discovery of EM afterglow 
proving definitive. (c) If three or more subthreshold events are detected by the 
clustering analysis (shown here for all possible combinations that include at 
least one event from a given stream), or if some other constraint is applied, 
the FPR is reduced to a level that enables a definitive or near-definitive claim 
of discovery. 
},
pos=here,
label=table.FPR
]
{cccccccccc}
{\tnote[\#]{Auger events for the pairwise (and optionally threefold) analysis are selected with 
galactic latitude within $\pm 5^\circ$ and energy $\ge$ 1 EeV.}
\tnote[$\dagger$]{Coincidence rate of event pairs with serendipitous 
follow-up detection of a GRB or SNe light curve (not including galactic searches).} 
\tnote[*]{Event pairs with above threshold EM detection. GBM not included as the high significance partner, due to poor spatial localization.}
\tnote[$\ddagger$]{An additional temporal cut of $\Delta T = 1$ s
is applied to the pairwise analysis for TeV and higher observatories, to search for 
primordial black holes and other exotic phenomena. Shown here are the FPRs for pairwise coincidences with HAWC.}
}
{
%
& & 	\begin{sideways} IceCube~ \end{sideways} 	&
         \begin{sideways} ANTARES\end{sideways} 		&
	\begin{sideways} LIGO-Virgo \end{sideways} 		&
	\begin{sideways} Auger \end{sideways} 		&
	\begin{sideways} BAT \end{sideways} 		&
	\begin{sideways} GBM \end{sideways} 		&
	\begin{sideways} LAT \end{sideways} 		&
	\begin{sideways} HAWC \end{sideways} 		\\
\cline{2-10}
\noalign{\smallskip}
\cline{2-10}
%
\multicolumn{1}{l }{(a) Single} 	& \multicolumn{1}{r |}{Above thresh.} 
& $\sim$0			& $\sim$0			& $\sim$0			&  $\sim$0		& $\sim$100		& $\sim$250		& $\sim$10		& $\sim$10			\\ 
\multicolumn{1}{l}{streams}	& \multicolumn{1}{r|}{Subthreshold} 
& 8.8$\times10^4$	& $2.9\times10^4$	& 3.2$\times10^3$ 	& 2.4$\times10^5$	& 1.4$\times10^5$	& 3.1$\times10^2$	& 3.9$\times10^4$	& 2.6$\times10^4$ 	\\ 
\cline{2-10}
%
\multicolumn{1}{l}{} 			& \multicolumn{1}{r|}{IceCube}		&30		&1.5		& 35		&1.8		& 11		&10		& 24		&6.5	\\
\multicolumn{1}{l}{} 			& \multicolumn{1}{r|}{ANTARES}	& 1.5		&0.5		& 12 		& 1.1		& 0.7		&3.5 		&7.1		&0.6	\\
\multicolumn{1}{c}{}  			& \multicolumn{1}{r|}{LIGO-Virgo}	& 35 		&12		& \na		& 8.4		&53		&0.6		&16		&10	\\
\multicolumn{1}{l}{(b) Pair-}& \multicolumn{1}{r|}{Auger\tmark[\#] }	& 1.8 	&1.1		& 8.4 	& 20		&2.9		& 2.5		&5.9		&1.5	\\
\multicolumn{1}{l}{wise FPR}   	& \multicolumn{1}{r|}{BAT} 		& 11 		&0.7		& 53 		& 2.9 	& \na		&16		& 32		&3.3	\\ 
\multicolumn{1}{c}{}		   	& \multicolumn{1}{r|}{GBM} 		& 10 		&3.5		& 0.6 	& 2.5 	& 16		&\na		& 5.0 	&3.2	\\ 
\multicolumn{1}{c}{}               	& \multicolumn{1}{r|}{LAT} 		& 24 		&7.1		& 16 		& 5.9 	& 32 		& 5.0 	& \na		&6.8	\\ 
\multicolumn{1}{c}{}             	& \multicolumn{1}{r|}{HAWC} 		& 6.5 	& 0.6		& 10 		& 1.5 	& 3.3 	&3.2 		& 6.8 	&\na\\ 
\cline{2-10}
\noalign{\smallskip}
\cline{2-10}
%
\multicolumn{1}{l}{} 			& \multicolumn{1}{r|}{GRB lt.~curve\tmark[$\dagger$]}	
& 0.071	&0.003	& 0.16	& -		& 0.0004	&0.08	&0.13	& 0.019	 	\\
\multicolumn{1}{l}{} 			& \multicolumn{1}{r|}{SNe lt.~curve\tmark[$\dagger$]}	
& 1.5		&0.07	& 3.4		& -		& 0.009	&1.6		& 2.7 	&0.4	 		\\
\multicolumn{1}{l}{(c) High} 	& \multicolumn{1}{r|}{3-fold coinc.} 					
& 0.15	& 0.03 	& 0.31	& 0.64 	& 0.12	& 0.09	& 0.40	& 0.08		\\
\multicolumn{1}{l}{significance}& \multicolumn{1}{r|}{3-fold coinc\tmark[\#]} 					
& 0.10	& 0.02 	& 0.15	& 0.06 	& 0.08	& 0.04	& 0.23	& 0.04		\\
\multicolumn{1}{l}{}			& \multicolumn{1}{r|}{High-sig.~EM\tmark[*]} 					
& 0.015	&0.002	& 0.045	& 0.044 	&0.010 	& 0.014 & 0.039 	& 0.005		\\
\multicolumn{1}{c}{}                   	& \multicolumn{1}{r|}{PBH search\tmark[$\ddagger$]} 	
& 0.13 	&0.01	& -		& 0.21	& - 		& - 		& - 		& 0.35		\\  
\cline{2-10}
\noalign{\smallskip}
\cline{2-10}
}

\end{landscape}

\section{Discovery Power of AMON}
\label{sec.power}

With the array of sources and messenger types described in
Sec.~\ref{sec.opp} and the possible groupings of observatories
described in Sec.~\ref{sec.ops}, there are a multitude of ways in
which AMON may provide scientific gain. Unable to cover all features
of AMON in one paper, we focus here on three examples.
Sec.~\ref{sec.nugamma} examines how the distribution of AMON Alerts
can greatly enhance the search for electromagnetic counterparts to
candidate sources of \HENlong, while Sec.~\ref{sec.gwhen} quantifies
the gains for a search for joint sources of gravitational waves and
\HEN. Sec.~\ref{sec.pbhexample} discusses an example search for exotic
phenomena, showing how an AMON analysis may provide the first clear
signature for primordial black holes.

\subsection{Follow-up of candidate \HENsing\ sources}
\label{sec.nugamma}
As a first example, we consider the follow-up imaging of candidate
\HENsing\ sources. Such searches are already underway via efforts
by individual neutrino observatories and their follow-up partners.  
For example, in \cite{Franckowiak:1204968}, the IceCube Observatory
identifies clusters of two or more neutrino events as a trigger for 
follow-up imaging. Due to the high rate of atmospheric neutrinos and 
other backgrounds, the expected FPR for a pair of up-going events for 
IceCube-86 is approximately 30 yr$^{-1}$ (Table~\ref{table.FPR}, row ``b''). 
Alerts are then distributed to the ROTSE telescope array or, after using a
maximum likelihood analysis to refine their number, to Swift and
PTF~\cite{2011ICRC....4..185H, 2011arXiv1111.2741T}. This program is
already producing the first limits on SNe with choked jets~\cite{2012A&A...539A..60A}.

As a shorthand, we introduce the notation ``$2\nu$'' to indicate a trigger 
condition of 2 or more coincident neutrinos, or $\nu$-$N\gamma$ for one or more neutrinos and $N$ or more 
{\gray}s. To be concrete, we study the 
follow-up imaging of $2\nu$ alerts at a rate of 10 yr$^{-1}$ using Swift's co-aligned X-ray and UV/Optical 
Telescopes (XRT, UVOT), with this considered a {\it status quo} approach for the scientific 
community.   Due to the narrow field of view ($0.4^\circ$ diameter) of the XRT and UVOT,
7 Swift pointings are required to cover the error region of each alert,
totaling  70 pointings in all.

In contrast, the AMON approach enables the realtime coincident analysis of
neutrino and electromagnetic data streams, comparing the arrival time
and direction of a {\it single} candidate \HEN\ with data from Swift's large field of 
view Burst Alert Telescope (BAT), Fermi's Large Area Telescope (LAT), or the 
upcoming ground-based HAWC.   
Because of the superior AMON localizations, derived via the joint analysis 
with high-energy $\gamma$-rays, the same follow-up telescope resource
(70 pointings) can be used for follow-up of a significantly-increased
number of AMON alerts. Here we consider a program that follows up
five $2\nu$ alerts (at 7 pointings each) and 35 $\nu$-$N\gamma$ alerts (1 pointing each).  
For the $\nu$-$N\gamma$ alerts, we considered the pairing of IceCube with both BAT and LAT.

\begin{SCfigure}[50][h]
\includegraphics[width=0.5\textwidth]{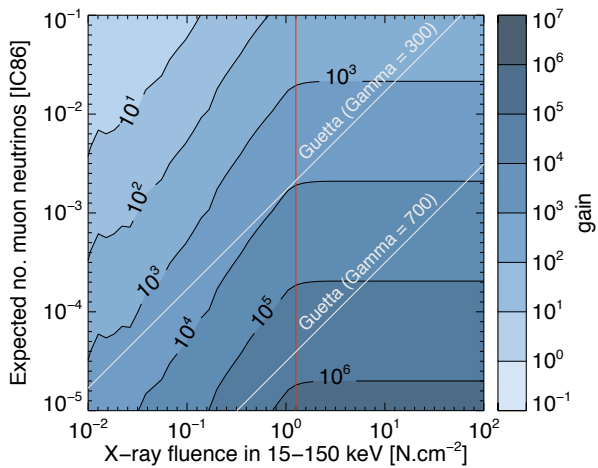}
\caption{%
  \small Sensitivity of the AMON multi-channel method compared to the
  {\it status quo}, presented as a ratio ranging up to $\sim10^6$, for
  follow-up observations of candidate astrophysical neutrino sources.
  The assumed shapes for both gamma and neutrino spectra follow a
  broken power law with initial and final logarithmic slopes of
  $\alpha = 1.0$ and $\beta =2.0$, and break energies
  $\mathcal{E}_\gamma = 0.2$ MeV and $\mathcal{E}_\nu = 0.35$ PeV.
  The vertical line shows a typical threshold for a hard \xray\
  observatory ({\it e.g.}, Swift BAT), and diagonal lines corresponding to
  predicted neutrino to gamma-ray fluxes under a range of current
  theoretical GRB models \cite{2004APh....20..429G}.  }
\label{fig.nugamma}
\end{SCfigure}

The relative sensitivity of the {\it status quo} and AMON approaches
was studied via a Monte Carlo calculation. As a source model, we assumed that both
\gray\ and neutrino spectra follow a broken power-law with initial and
final logarithmic slopes of $\alpha = 1.0$ and $\beta =2.0$, and break
energies $\mathcal{E}_\gamma = 0.2$ MeV and $\mathcal{E}_\nu = 0.35$
PeV, respectively. The neutrino and \gray\ fluence was allowed to vary, as indicated
by the axes of Fig.~\ref{fig.nugamma}.
Using realistic models for the IceCube-86, BAT, and LAT effective
areas, point spread functions, and their overlapping field of views,
we were able to determine the acceptance of each simulated
source. Realistic background models were adopted for each observatory,
enabling the calculation of false positive rates for each joint
analysis.

In each case, the FPR was tuned via a likelihood analysis of the relative position of the $\nu$ and
$\gamma$ events. Subthreshold BAT and LAT events were allowed into the simulated data stream, 
with terms added to the likelihood function to favor the more significant EM signals. Furthermore, we added
a regulating term to the likelihood function to account for the possibility of a real \gray\ source in accidental 
coincidence with an IceCube background event ({\it e.g.} an atmospheric neutrino). Once the FPR is chosen for 
any pair of observatories, we identify the required likelihood threshold, and can then calculate the corresponding
sensitivity to our modeled source.   While the initial FPR is quite high (10-70 alerts yr$^{-1}$), the final FPR
will be orders of magnitude lower after the inclusion of follow-up data and, as such, can be ignored in the study of 
sensitivity. While here we have characterized the sensitivity of the triggering methods (AMON versus {\it status quo}),  
there are a number of unknown factors in the sensitivity of the subsequent follow-up searches; for example, due to 
optical magnitude. However, we study the ratio of the sensitivity of the two methods, expecting the acceptance of
the two follow-up programs will be approximately the same and to cancel in the ratio.

%
%

The results are shown in Fig.~\ref{fig.nugamma}, where substantial gains in sensitivity (up to $10^6$) are realized.
The sensitivity of each method is defined as the fraction of true
sources pursued using a given follow-up capability. 
In addition to enabling pursuit of an increased number of alerts
overall, the AMON approach realizes these large sensitivity
gains 
by triggering from {\it single} neutrinos rather than a neutrino pair,
leveraging the increased interaction strength of $\gamma$'s over
$\nu$'s. The left axis of Fig.~\ref{fig.nugamma} is defined in such a way 
that it could equally well describe the expected event rate for ANTARES, 
with only a small change in the contours due to the narrower 
neutrino point spread function.

Gains of this type have also been demonstrated for gravitational wave
source searches, using GW-$\gamma$ triggering over the {\it status
  quo} GW-only methods~\cite{2008CQGra..25k4051A}. For GW-only
searches, the poor GW localization is filtered by requiring
coincidence between the GW localization and a nearby cataloged
galaxy~\cite{2012arXiv1205.1124E, 2012A&A...539A.124L}.  While useful,
this approach suffers from several shortcomings: First, it depends on
the completeness and astrophysical relevance of the associated galaxy
catalog; second, it cannot be extended to the subthreshold regime due
to the dramatic increase in false alarm rates; and finally, the
approach is limited to an horizon of $\simlt$100 Mpc, beyond which
galaxy fields become crowded and catalogs are highly incomplete. The
multi-channel, multimessenger approach by contrast provides a unique
candidate source position for follow-up pointing, and leverages the
coincidence technique to push into the subthreshold, possibly
revealing otherwise-undetectable source populations.  With Advanced
LIGO expected to ultimately achieve sensitivity out to
$\simgt$500\,Mpc~\cite{2010CQGra..27q3001A}, the AMON approach can offer 
a $>$125-fold gain in sensitivity for events with bright EM counterparts, or 
$>$1000-fold for regions that have the least complete galaxy catalog.

\subsection{Joint detection of neutrinos and gravitational waves}
\label{sec.gwhen}

Here we consider the joint detection of gravitational waves and
high energy \HENlong\ as an example of a science investigation
enabled by the realtime operation of AMON. The search for jointly
emitting GW+\HEN\ sources has been addressed in
\cite{2011PhRvL.107y1101B, 2012JPhCS.363a2022B}, including predictions
for Advanced LIGO-Virgo and IceCube-86, and we follow the same general
approach (see also \cite{2008CQGra..25k4039A, 2012arXiv1205.3018T}).  
However, while previous authors focus on the joint analysis
of a multimessenger data archive, we emphasize the possibility of
carrying out this analysis in realtime.
We characterize the sensitivity of a blind (untriggered) search by
Advanced LIGO-Virgo using the horizon distance $\DGW$, inside which a
GW source is expect to generate an above-threshold detection,
\begin{equation}
\DGW \approx 80~\text{Mpc} \left(\frac{\EGW}{10^{-2} \Msun}\right)^{\frac{1}{2}}
\label{eqn.DGW}
\end{equation}
where $\EGW$ is the isotropic equivalent gravitational wave energy of
the source. 
This is based on the results of \cite{2010PhRvD..81j2001A} and assumes 
little of the GW waveforms.
Without AMON-like infrastructure, a real-time detection of
both GW and \HEN~requires that a source be within this horizon
{\it and} that it produce a neutrino flux sufficient to trigger $n_\nu
\ge 3$ events in IceCube or ANTARES (since $n_\nu \le 2$ would lead to
an inconclusive false positive rate).  These detections would then be compared {\it
  post facto}, either through an archival analysis or {\it
  above-threshold} alert distribution ({\it e.g.}~through GCN), to determine
if there are any joint GW-\HEN~sources\footnote{In
  \cite{2011PhRvL.107y1101B}, the authors consider a slightly
  different problem, being interested in the above-threshold detection
  of GW {\it or} \HEN. While an interesting measure, their
  calculations do not provide the sensitivity for above-threshold
  detection of joint sources that we need for comparison to AMON.}.

By contrast, a system like AMON will provide two-way communication
between observatories, enabling a realtime triggered search of GW
data, either by transmission of pre-alerts from AMON to LIGO-Virgo or
by carrying out a triggered search on subthreshold GW data transmitted
to AMON (albeit in some preprocessed form). Searches of this kind suffer
a lower trials penalty and may assume various templates for the
sources. For example, the postulate that \HEN\ are detected in
coincidence allows one to make an assumption about the orientation of
the source, making the GW signal more favorable for detection
\cite{2012arXiv1206.0703C}. In all, we estimate that these effects
will combine to increase the horizon distance to $2 \times \DGW$~\cite{2012arXiv1206.0703C}.
Furthermore, only a single neutrino event is
required to trigger the analysis, providing significant gain for
this approach since the expected number of neutrinos has been
observed to be typically $\ll 1$~\cite{2012Natur.484..351A}.

Since we are considering the relatively deep horizon distance for
Advance LIGO-Virgo (so that galaxy distribution can be assumed
uniform), but not so far that redshift effects become important, it is
sensible to apply the effective volume approach of \ref{sec.Pdv},
integrating up to the physical limit of $\DGW$ or $2\DGW$.  The
resulting effective volumes are given in Table~\ref{table.Veff2}.

\ctable[ caption={\small Effective volumes for a joint GW-\HEN~search, 
assuming $\EGW = 10^{-2} \Msun$ and $\EHEN = 10^{49}$ erg
  for a choked GRB model.}, pos=here, label=table.Veff2 ] {cccc}
{
  } { \hline \hline
  Analysis 						& Integration 					& Neutrino 		&$\Vav$  \\
  & limit (Mpc) 					& threshold 		&  [Mpc$^{3}$] \\
  \hline
  GW+\HEN				& $80$ 						& $n_\nu \ge 3$	& $4.6\times10^2$ \\
  AMON						& $160$						& $n_\nu\ge 1$		& $5.9\times10^4$\\
  \hline }

Under the assumption of a low false positive rate, the 90\% upper
limit (chosen for consistency with \cite{2011PhRvL.107y1101B}) for the
joint GW-\HEN~source density is given by
\begin{equation}
\rgwnu < \frac{2.3f_b}{T\Vav}~~ [\text{yr}^{-1}\text{Mpc}^{-3}] 
\end{equation} 
where $T$ is the livetime, which we take to be one year, $\Vav$ is the
solid angle averaged effective volume for the given method, and $f_b$
is the beaming factor for neutrinos. We take the latter to be $f_b =
14$ (from \cite{2011PhRvL.107y1101B}) but it will cancel in our final
result.

The effective volume depends on $\EGW$ through equation
\ref{eqn.DGW}. For neutrinos, the expected number of detections from a
fiducial source at distance $r$ can be related to the neutrino
isotropic energy $\EHEN$, using a linear relationship provided in
\cite{2011PhRvL.107y1101B},
\begin{equation}
n_\nu \approx \frac{1}{\kappa} \left(\frac{\EHEN}{10^{49}\text{erg}}\right)\left(\frac{10~\text{Mpc}}{r}\right)^2
\end{equation}
where, for IceCube-86, $\kappa = 0.75$ for a high luminosity GRB model and $\kappa = 1.6$ for a model of choked GRBs.  

\begin{figure}[h]
\includegraphics[width=0.5\textwidth]{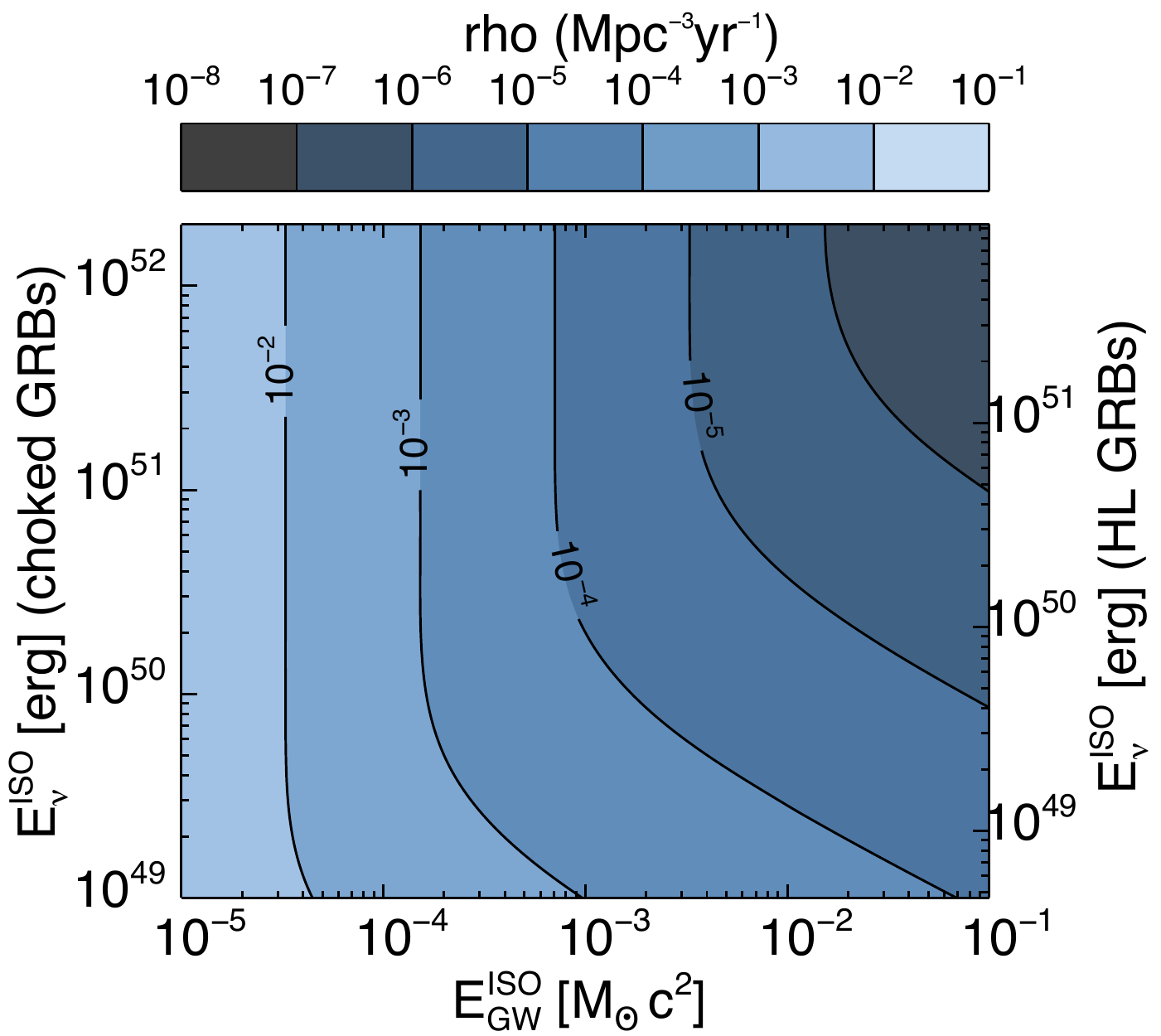}
\includegraphics[width=0.5\textwidth]{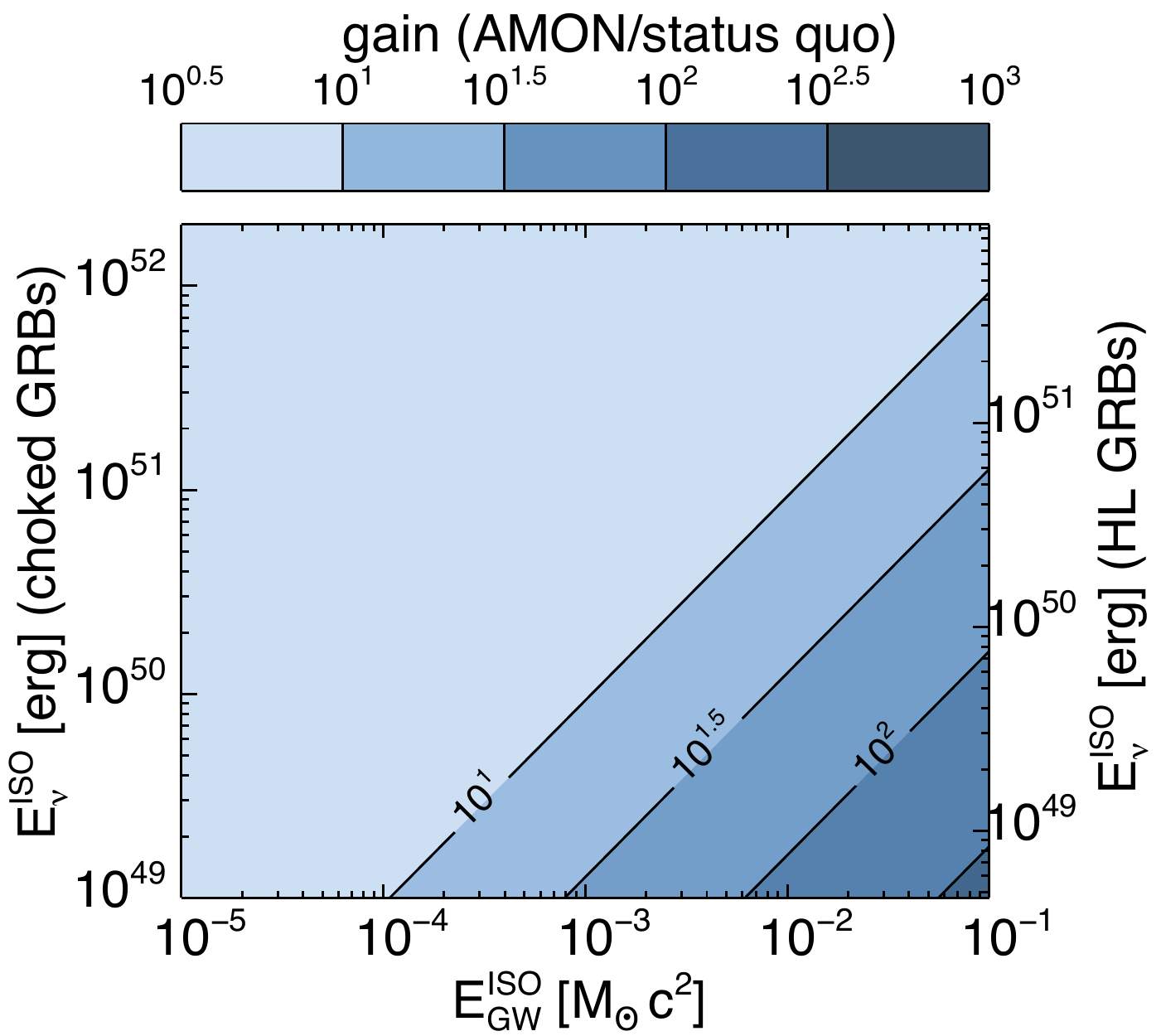}
\caption{\small (Left) Projected GW+\HEN~source population upper limits
  after one-year of AMON utilizing observations with Advanced
  LIGO-Virgo and IceCube-86. (Right) Ratio of the sensitivity for the
  AMON approach to the sensitivity of the status quo approach (as
  described in the text).  }
\label{fig.gwhen}
\end{figure}

The results are shown in Fig.~\ref{fig.gwhen}. The left side shows the
upper limit in units of $\text{yr}^{-1}\text{Mpc}^{-3}$ that would be
obtained by a non-detection with AMON. The right side shows the
multiplicative gain in sensitivity that the AMON approach provides
over an analysis that requires above-threshold detection in both GW
and \HEN~channels. The gain is most significant for bright GW sources
and weak neutrino sources, where the burden for discovery is shifted
from the \HEN~to GW channel. However, the gain is strictly greater
than 1, with an asymptote of $8$, representing the increase in
effective volume due to triggering the GW search.

\subsection{Searching for primordial black holes}
\label{sec.pbhexample}

Here we consider the search for primordial black holes as an
example of a collaborative study of accumulated multi-facility data,
as enabled by the AMON infrastructure and data sharing policies.  In
particular, we have modeled a joint search with the IceCube neutrino
observatory, HAWC TeV gamma-ray observatory, and the Pierre Auger
cosmic ray observatory. All are sensitive to particles produced in the
final stages of PBH evaporation, including a potential signal from
ultrahigh-energy neutrons detected at Pierre Auger, since these
neutrons would not suffer from the magnetic deflection and time delay
effects of charged cosmic rays (the chief UHE neutron background).

Based on the energy threshold of these experiments and energy
dependence of various models predicting energy spectra of PBHs in
their final stage of evaporations ({\it e.g.}~\cite{MCGIB1, HECKLER,
  KAPUSTA}), the model of a non-rotating, uncharged black hole without a
chromosphere by MacGibbon and Webber~\cite{MCGIB1} was chosen for
estimation of the expected signal at each of these detectors. The time
integrated particle spectra above 100 GeV were computed following
methods from refs.~\cite{HAL1, MCGIB1, MCGIB2, HAL2}.  The main
particle decay chains considered in our calculations are described in
\ref{sec.pbh}.

\begin{SCfigure}[50][!bth]
\includegraphics[width=0.43\textwidth]{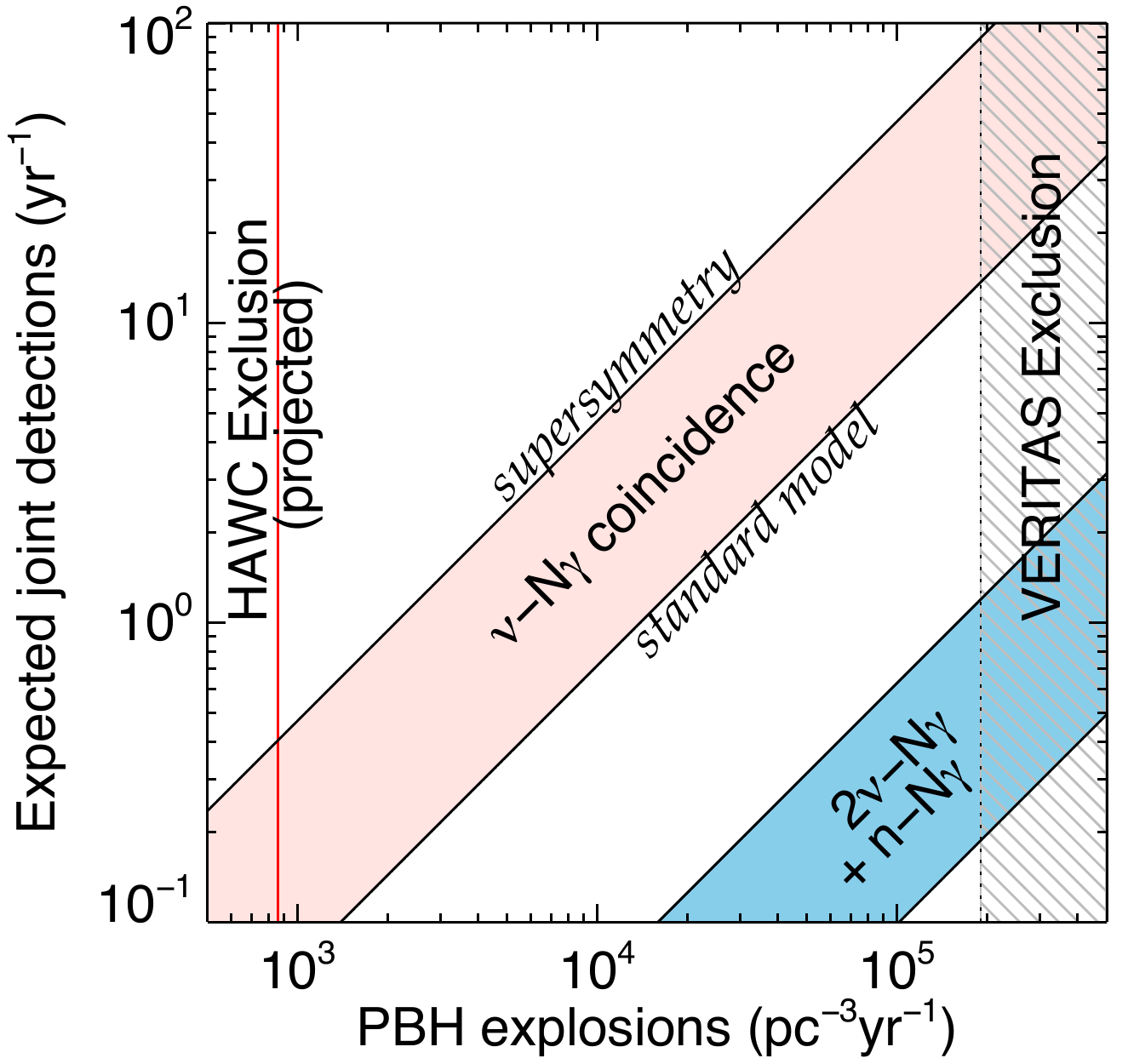}
\caption{Expected number of sources detected in at least two of
  the $\nu$, $\gamma$, and $n$ channels, modeled here with the
  IceCube-86, projected HAWC, and Auger sensitivity, as a function of
  the primordial black hole density. The shaded region to the
  right indicates PBH densities that previous direct searches have
  excluded, while the vertical line shows where the upper limit is
  expected to move if HAWC does not observe a gamma ray excess after a
  year of livetime. In between, there is the possibility of positive
  joint detections, with potentially a unique PBH signature.  }
\label{fig.pbh}
\end{SCfigure}

It is anticipated that, when fully operational, HAWC by itself will
either detect or provide the best upper limit on direct searches for
PBH decay, around 2 orders of magnitude better than the current upper
limit (as shown in Fig.~\ref{fig.pbh}). 
%
%
%
%
To do so, HAWC must perform a blind search of its own data. After a
cut to remove many hadronic events, 5.1 kHz of subthreshold trigger
events remain, which will be analyzed for coincidences across a
spatial template with $10^4$ trials \cite{2012APh....35..641A}. Taking
a generous time window of $\Delta T = 5$ s and a number threshold of
$n \ge 20$, the HAWC false coincidence rate can be reduced to $< 1$
yr$^{-1}$. However, it is expected that HAWC will also observe real
astrophysical transients that are unrelated to PBHs, forming a second
type of false positive for the PBH search. We conservatively estimate
this number to be $\FPRname{HAWC} < 17$ bursts per year, approximately
equal to the known LAT GRB rate, allowed to fluctuate up by
$2\sigma$. If, after $T=1$ year of livetime, $17$ unidentified bursts
were observed, the 99.73\% ($3 \sigma$) upper limit for PBH would be
given by
\begin{equation}
 \rpbh < \frac{32.1}{\Vav}~~\left[\text{yr}^{-1}\text{pc}^{-3}\right]
\end{equation} 
where 32.1 is chosen to give the 99.73\% upper limit of the Poisson
distribution, and $\Vav$ is the integral of the trigger condition $P(n
\ge 20 | r)$ over $4\pi r^2dr$, averaged over solid angle, where $r$
is the distance from the Earth.  The value of $\Vav$ is specific to
the sensitivity of the observatory and the intrinsic $\gamma$-ray
fluence of the fiducial source, and is further described in
\ref{sec.Pdv}. For the current calculation, we estimate $\Vav = 0.037$
pc$^{-3}$. The calculated value of $\rpbh <
3.5\times10^3~\text{yr}^{-1}\text{pc}^{-3}$ can be compared to the
current best limit of $1.9\times10^5~\text{yr}^{-1}\text{pc}^{-3}$
from VERITAS~\cite{VERITAS_PBH}.
    
Between the current VERITAS limit and the projected HAWC limit, there
exists the possibility of a future positive detection of PBH
decays. However, the HAWC data by itself may be unable to distinguish
between PBH and other scenarios. A joint AMON study of HAWC data with
neutrino and neutron data (modeled here with IceCube and Auger), will
enable the search for multimessenger coincidences, providing a unique
PBH signature. Just a few such coincidences should reveal a timing and
energy structure that are indicative of Hawking evaporation.

\ctable[caption={\small Effective volumes for multimessenger PBH
  search. Values are averaged over $4\pi$ sr, with only those regions
  of overlapping sensitivity contributing to each pairwise
  calculation.},pos=!tbh,label=table.Veff] {cccccc} {} { \hline
  \hline
  \multicolumn{2}{c}{Observatories} &\multicolumn{2}{c}{Trigger conditions}  &$\Vav_{SM}$ & $\Vav_{SUSY}$  \\
  A & B & A & B & [pc$^{3}$]  & [pc$^{3}$]\\
  \hline
  HAWC	& -	& $n_\gamma \ge 20$ & - & 0.0374 & 0.245\\
  IceCube & HAWC & $n_\nu \ge 1~$ & $n_\gamma \ge 13$ & 9.8$\times 10^{-5}$ & 6.4$\times 10^{-4}$\\
    IceCube & HAWC & $n_\nu \ge 2~$ & $n_\gamma \ge 13$ & 1.3$\times 10^{-6}$ &  8.3$\times 10^{-6}$\\
  HAWC & Auger & $n_\gamma \ge 13$ & $n_n \ge 1~$ & 1.3$\times 10^{-7}$ &  8.6$\times 10^{-7}$\\
  Auger & IceCube & $n_{n} \ge 1~$ & $n_\nu \ge 1~$ & 4.9$\times 10^{-9}$ &  3.2$\times 10^{-8}$\\
  \hline }

Folding together the PBH source model from \ref{sec.pbh} and the
volume integrals of \ref{sec.Pdv}, we calculate the effective volumes
given in Table~\ref{table.Veff}.
To maximize sensitivity to neutrinos and neutrons, we use number
thresholds of $n_\nu\ge 1$ and $n_n \ge 1$. For $\gamma$-rays, the low
false alarm rate can be relaxed significantly by requiring coincidence
with a single $\nu$ or $n$, allowing the HAWC threshold to be lowered
to $n_\gamma \ge 13$.
The expected rate of {\it true positive} joint detections in channels
$a$ and $b$ is then given by
\begin{equation}
\TPR{ab} = \frac{\rpbh}{\Vav_{ab}} ~~\left[\text{yr}^{-1}\text{pc}^{-3}\right]
\end{equation}
which we show as a function of $\rpbh$ in Fig.~\ref{fig.pbh}. Over the
range of $\rho$ which might lead to future positive detections by
HAWC, the total number of expected multimessenger coincidences ranges
from 0.1 to 100 yr$^{-1}$. Most of these are $\nu$-$N\gamma$ ({\it
  i.e.}~a coincidence between a single neutrino and $N$
$\gamma$-rays). However, there is a non-negligible possibility for
observing $n$-$N\gamma$ coincidences (where the neutron is detected by
Auger) or $2\nu$-$N\gamma$, providing an exciting possibility for
detecting three subthreshold events in coincidence (where the
$N\gamma$ from HAWC are treated as a single subthreshold event).

\section{Summary and Conclusions}
\label{sec.conclude}

The Astrophysical Multimessenger Observatory Network under development
at Penn State will link multiple high-energy, multimessenger, and
follow-up observatories together into a single comprehensive
system. We have explored the scientific opportunity for AMON
(\S\ref{sec.opp}), which is centered on the discovery and exploitation
of multimessenger transients. The nature of the brightest such
transients, which may manifest as the first detections of
gravitational waves, high-energy cosmogenic neutrinos, or high-energy
cosmic ray neutrons, is still unknown; likely candidates include
blazar flares and a diverse array of $\gray$ bursts and supernovae, as
well as exotic phenomena such as the evaporation of primordial black
holes from the early universe.

We have described the design, infrastructure, and current and
projected partner facilities of AMON (\S\ref{sec.ops}), showing that
the wide fields of view, high duty cycles, and subthreshold event
rates of the facilities are such that a robust and automated
statistical search for coincident events seen in the data of two or
more facilities is both interesting and feasible
(Table~\ref{table.FPR}). In particular, false positive rates for the
resulting AMON Alerts are sufficiently low that comprehensive
ground-based optical follow-up campaigns can be contemplated. 

With neutrino and \gray\ observatories are already participating in
AMON, and discussions underway with cosmic ray and gravitational
wave observatories, as well as multiple follow-up observatories, AMON
is poised to begin real-time operations within a year. 

To demonstrate the power of the AMON approach, we have carried out
three sets of simulations using current theoretical models of
multimessenger phenomena (\S\ref{sec.power}). In the first simulation,
we explored the challenge of EM counterpart searches for candidate
GRB-associated \HEN\ observed in IceCube, showing that vetting
candidate \HEN\ events against multiple EM data streams realizes a
$>$1000-fold gain in the efficiency of EM follow-up observations
(Fig.~\ref{fig.nugamma}). In the second simulation, we explored the
improvement in search sensitivity for jointly emitting
GW+\HEN\ transients that is realized by extending these searches into
the subthreshold regime, rather than restricting the search to events
that generate statistically-significant signals in both channels; we
find that a $>$10-fold increase in event rates, or improvement in
upper limits, is easily achieved (Fig.~\ref{fig.gwhen}). In the third
simulation, we explored the multimessenger signature of PBH
evaporation, demonstrating that -- consistent with current upper
limits on the local PBH density -- the coincidence of an
IceCube-detected neutrino with a cluster of HAWC-detected
$\gamma$-rays could provide evidence for observation of a PBH
evaporation event within the first year of HAWC operations
(Fig.~\ref{fig.pbh}).

In addition to carrying out the real-time multi-facility transient
searches needed to realize these and other science gains, and enabling
follow-up EM observations by distributing transient alerts to
interested observers, AMON will provide a framework for large
observatory collaborations to work together on targeted archival
analyses using AMON's comprehensive events database and coincidence
analysis toolkit.
In these several ways, AMON will leverage and enhance the capabilities
of existing and future high-energy and multimessenger observatories,
powering a new and ambitious exploration of the transient universe
using all four forces, and so helping to realize the immense
promise of this dawning age of multimessenger astronomy.

\section*{Acknowledgements}

Initial development of AMON has been funded by Penn State's Office of
the Senior Vice President for Research, the Eberly College of
Science, and the Penn State Institute for Gravitation and the Cosmos.
D.~F.~Cowen acknowledges the support of the Penn State Institute for
CyberScience Faculty Fellows Program; I.~Bartos and S.~M\'arka
acknowledge support from Columbia University and the National Science
Foundation under cooperative agreement PHY-0847182. 

The authors acknowledge valuable discussions and inputs from J\"urgen
Brunner, James Chiang, Alessandra Corsi, Farhan Feroz, Anna
Franckowiak, Gabriela Gonzalez, Jordan Goodman, Kazumi Kashiyama, Jim
Matthews, Julie McEnery, David Miller, Prasenjit Mitra, Tom Prince,
Padma Raghavan, Soeb Razzaque, Ben Whelan, and Weikang Zheng.

\clearpage
\appendix

\section{Effective Volume Integrals}
\label{sec.Pdv}
In a number of examples in the text, a simple population model was applied that assumes a uniform spatial distribution of sources with a fiducial set of source parameters (encapsulated here by the parameter $\lambda$). In the case where  $z \ll 1$, we can define an effective volume $\Veff$  for a specific set of trigger conditions by the Euclidean volume integral,  
\begin{equation}
\frac{d\Veff}{d\Omega} =   \frac{1}{4\pi}\int_0^{D} 4\pi r^2 dr P(\text{true} |\lambda,  r),
\end{equation}
where $P(\text{true} | r, \lambda)$ is the conditional probability of a true positive detection, given the fiducial source parameters and distance to the source $r$. The maximum distance $D$ may represent either a physical limitation of the method or else a parameter that we use to regulate the integral before taking $D\to\infty$.  The parameter(s) $\lambda$ may depend on pointing direction, and so we have written $\Veff$ as differential with respect to solid angle $\Omega$. 
The expected rate of source detections is then given, after integrating over solid angle, by
\begin{equation}
R = \rho \Veff,  
\end{equation}
where $\rho$ is the intrinsic rate per unit volume of transient sources. In the special case where the observatory is equally sensitive across its field of view, $\frac{d\Veff}{d\Omega}$ is independent of pointing and we can write
\begin{equation}
R = \rho \Omega\frac{d\Veff}{d\Omega}.
\label{eqn.Veff_flat}
\end{equation}

In the case where the participating observatories are each monitoring for Poisson processes, $P$ can be taken to be the cumulative Poisson probability distribution above some set of number thresholds $\vec n_0$. The vector notation includes the possibility of studying events from multiple observatories, with indices $a = 1\to M$, so that
\begin{equation}
P(\vec n \ge \vec n_0 | \vec\lambda, r) = \prod_{a=1}^{M}\left( 1- \sum_{k_a = 0 }^{n_{0a}} \frac{ e^{-\lambda_a/r^2}}{k_a!}
\left(\frac{\lambda_a}{r^2}\right)^{k_a}\right), 
\end{equation} 
where $\lambda_a \equiv \frac{1}{4 \pi} \int dE \frac{dN_a}{dE} \Aeff_a(E)$ is a measure of the number of particles that can be detected by observatory $a$ (where $N_a$ is the total number of particles produced and $\Aeff_a$ is the observatory's effective area). It is normalized in such a way that $\lambda_a/r^2$ is the expected number of detections for a source at a distance $r$.   The trigger condition $\vec n \ge \vec n_0$ is shorthand for simultaneous  detections above the specified number threshold for each observatory. 

As such, we are interested in integrals of the form
\begin{eqnarray}
v_{\vec k}(\vec{\lambda}) &\equiv&  \int_0^{D}  r^2 dr \prod_{a=1}^{M}\frac{ e^{-\lambda_a/r^2}}{k_a!}
\left(\frac{\lambda_a}{r^2}\right)^{k_a} \nonumber \\
&=& \left( \prod_{a=1}^{M}  \frac{\lambda_a^{k_a}}{k_a!}    \right)  \int_{\Lambda/D^2}^{\infty} dx \frac{\Lambda^{\frac{3}{2}} e^{-x} x^{\kappa-\frac{5}{2}}}{2\Lambda^{\kappa}}\nonumber\\
&=&   \frac{\Lambda^{\frac{3}{2}}}{2\kappa!}Q_{\vec{k}}(\vec{\lambda})~\Gamma\left(\kappa - \frac{3}{2}, \frac{\Lambda}{D^2}\right),
 \label{eqn.v2d}
\end{eqnarray}
where we have changed the integration variable to $x = \Lambda/r^2$ and defined $\Lambda = \sum_{a} \lambda_a$,  $\kappa = \sum_{a} k_a$, and the term 
$Q_{\vec{k}}(\vec{\lambda})= \left(\frac{\kappa}{k_1,k_2,...}\right) \prod_{a} \left(\frac{\lambda_a}{\Lambda}\right)^{k_a}$. 
By the multinomial theorem,  we note that
\begin{eqnarray}
v_{\kappa}(\Lambda) &\equiv& \sum_{\sum k_a = \kappa} v_{\vec k}(\vec{\lambda}) \nonumber\\
&=& \frac{\Lambda^{\frac{3}{2}}}{2\kappa!}~\Gamma\left(\kappa - \frac{3}{2}, \frac{\Lambda}{D^2}\right),
\label{eqn.V}
\end{eqnarray}
where we have used a single subscript on the left hand side (rather than a vector). As expected, this is the same result as equation \ref{eqn.v2d} applied to a single observatory.  
The interpretation is that equation \ref{eqn.V} can be applied to triggers from multiple observatories that are interchangeable ({\it e.g.}~when considering $\kappa=2$ neutrinos form IceCube or ANTARES, regardless of which of the observatories triggered).  

Importantly, $v_{0}$ and $v_{1}$ diverge as $D \to \infty$, but the linear combination $\frac{D^3}{3} - v_0 - v_1$ is finite. Thus, if we are looking for a single Poisson trigger, then a physical limit $D$ must be applied to achieve a finite result ({\it e.g.}~by requiring that a single detected neutrino originate from within the sensing region of a gravitational wave network). However, if we instead require that $\kappa \ge 2$, then the result is finite as $D\to\infty$. 

We can also marginalize over one observatory, defining $\vec  k = (k_1, \vec{k}')$ and using the normalization of the Poisson series to find  
\begin{equation}
\sum_{k_1 = 0}^{\infty} v_{(k_1, \vec{k}')}(\lambda_1, \vec\lambda') =  v_{\vec{k}'}(\vec\lambda').
\end{equation}
We can apply this result to two observatories. In the case where we search for at least $m$ events from the first observatory and at least $n$ events from the second,
\begin{eqnarray}
\frac{d\Veff_{mn}}{d\Omega} &=& \frac{D^3 }{3} -  \sum_{ i, j\ge m,n}v_{(i,j)}(\lambda_1, \lambda_2) \nonumber \\
&=& \frac{D^3}{3} -  \sum_{i=0}^{m-1}v_i(\lambda_1)  -  \sum_{j=0}^{n-1}v_j(\lambda_2)  + \sum_{i=0}^{m-1}\sum_{j=0}^{n-1}v_{(i,j)}(\lambda_1, \lambda_2).
\label{eqn.Veff1}
\end{eqnarray}
However, if  we search for a total of $n$ events, regardless of which observatory they come from,  
\begin{eqnarray}
\frac{d\Veff_n}{d\Omega} 
&=& \frac{D^3}{3}  -  \sum_{j=0}^{n-1}v_j(\Lambda).
\label{eqn.Veff2}
\end{eqnarray}
Both equations \ref{eqn.Veff1} and \ref{eqn.Veff2} lead to finite results as $D \to \infty$.

We may wish to consider the case where one or both observatories have an effective area that changes significantly off-axis, meaning that $\frac{d\Veff}{d\Omega}$ depends on direction. In general, one must carry out the integration over solid angle numerically, although we have already noted the trivial analytic result in the case of equation \ref{eqn.Veff_flat}. In addition, there is an analytic solution in the case where only the first observatory has a dependence on zenith angle $\theta_z$, that dependence is proportional to $\cos\theta_{z}$, and the number threshold of the first observatory is $n_1 \ge 1$. However, the result is sufficiently complicated that we choose to omit it here. 


\section{Primordial Black Hole Source Model}
\label{sec.pbh}
Here we describe the main decay chains leading to the final particle spectra from the PBH explosion under the assumptions of the Standard Model (SM) and Supersymmetry (SUSY). 

\subsection{SM decay chains}
 During the final stage of the PBH life,  all of the 118 SM particle degrees of freedom ({\it dof}) are radiated away. Since there are more {\it dof} for quark and gluons  (72) than for leptons and photons (26), the final spectra of both HE and UHE gamma rays and neutrinos originate mostly from the decaying hadrons~\cite{MCGIB1}.  Neutrons and antineutrons are produced from these decaying quark fragmentation products, as well. 
 
We assumed that quark and gluon jets emitted during the final stage of PBH evaporation fragment into pions ($\pi^0$, $\pi^-$ and $\pi^+$) and baryons ($p$, $\bar{p}$, $n$, and $\bar{n}$) with branching ratios of 0.97 and 0.03~\cite{FRAG}, respectively. A fragmentation function  $\frac {\rm{d} N_{X}}{\rm{d}x}$ of quarks or gluons into a particle of type X has a form~\cite{FRAG}:
\begin{equation}
 \frac {{\rm d} N_{X}}{{\rm d}x}=\frac{15}{16}x^{-1.5}(1-x)^{2},
 \label{eg_f}
\end{equation}
where $x=E_{X}/E$, $E_{X}$ is the energy of a particle of type $X$, and $E$ is the quark/gluon jet energy. 

For baryon spectra, we assumed that equal numbers of $p$, $\bar{p}$, $n$, and $\bar{n}$ are produced after hadronization. Thus, the final $n$+$\bar{n}$ spectrum is obtained by multiplying distribution  $\frac {\rm{d} N_{X}}{\rm{d}x}$ by 0.015, convolving it with the Hawking primary spectra for quarks and gluons and integrating over time. 

The photon spectrum is obtained by convolving a flat photon energy distribution from $\pi^0$ decay with the 2$\cdot$1/3$\cdot$0.97$\cdot$ $\frac {\rm{d} N_{X}}{\rm{d}x}$ (where factor 2 accounts for the number of photons produced from each decaying pion and 1/3 is a fraction of neutral pions to the total number of pions) and the Hawking primary spectra. 

From decaying $\pi^-$ and $\pi^+$, $\bar{\nu}_{\mu}$ and $\nu_{\mu}$ are produced, together with $\mu^{-}$ and $\mu^{+}$. These neutrinos have flat energy distribution, but there are also secondary neutrinos originating from the consecutive muon decays: $\mu^{+(-)}\rightarrow e^{+(-)}+\nu_{e}(\bar{\nu_{e}}) + \bar{\nu_{\mu}}({\nu_{\mu}})$. Energy distributions of each of these neutrinos are convolved with 2/3$\cdot$0.97$\cdot$ $\frac {\rm{d} N_{X}}{\rm{d}x}$ and the quark/gluon Hawking spectra, and summed afterwards. Since production of $\nu_{\tau}$ from decaying hadrons is highly suppressed, most of this neutrino flavour originate from the direct Hawking radiation and from the decays of the directly emitted $\tau$ leptons. Therefore, one would expect a flux of $\nu_\tau$ that is almost two orders of magnitude lower than the total flux of other neutrino flavors. We neglected these two $\nu_\tau$ contributions to the total neutrino flux in this work. 

\subsection{SUSY decay chains}

If the Minimal Supersymmetric Standard Model (MSSM) describes Nature at the high-energy scales associated with the explosions of PBHs, the  number of {\it dof} available to be radiated increases by more than a factor of two:  244 compared to 118 {\it dof} in the SM, where we neglected graviton and gravitino {\it dof} and included five physical Higgs fields (see, for example, Ref.~\cite{MSSM}). This leads to an increase of the factor $\alpha(M)$ that directly affects the rate of evaporation by a factor of $\sim3.3$ when the PBH temperature reaches the SUSY particle production energy scale. From that point, the time left until complete evaporation will be shorter compared to the time predicted under the SM assumption. For example, one second before complete evaporation, the temperature of the MSSM PBH would be $T\sim 5.5$ TeV with almost 50\% more mass to be radiated than in the case of the SM PBH with $T\sim 8$TeV at the same time left before its complete evaporation. 

Given a number of unknown supersymmetric parameters (more than 100), for simplicity, we assumed that gluinos ($\tilde{g}$) are heavier than squarks ($\tilde{q}$) (as in, for example, the SPS1a benchmark scenario~\cite{SPA1} and the mSUGRA B benchmark model~\cite{MSUGRA}), thus they decay into antisquark/quark and squark/antiquark pairs~\cite{MSSM2}:
\begin{equation}
\label{eq_gq}
\tilde{g} \rightarrow \bar{\tilde{q}}+q, \tilde{q} + \bar{q}.
\end{equation}
The right chiral states of squarks would decay then mostly into the lightest netralino ($\tilde{\chi}^{0}_{1}$):
\begin{equation}
\tilde{q}_{R} \rightarrow \tilde{\chi}^{0}_{1}+q,
\label{eg_rd}
\end{equation}
whereas the left handed states would decay into charginos ($\tilde{\chi}^{\pm}_{1}$) or heavier neutralinos ($\tilde{\chi}^{0}_{2}$)~\cite{SQUARK}:
\begin{equation}
\tilde{q}_{L} \rightarrow \tilde{\chi}^{\pm}_{1}+q, ~~\tilde{\chi}^{0}_{2}+q.
\label{eg_ld}
\end{equation}

Further down these decay chains, we assumed that $\tilde{\chi}^{\pm}_{1}$ and $\tilde{\chi}^{0}_{2}$ preferably decay into leptons and charged leptons plus the lightest neutralino, respectively:
\begin{equation}
 \tilde{\chi}^{\pm}_{1} \rightarrow l^{\pm}\nu , ~~~~~~~~\tilde{\chi}^{0}_{2}\rightarrow l^{+}l^{-}\tilde{\chi}^{0}_{1}.
\label{eg_nd}
\end{equation}

Without carrying a complete calculation, our rough estimate for the multimessenger particle fluxes under the assumption of the MSSM are: $\sim$4 of the SM flux for neutrinos and $\sim$3.5 of the SM flux for photons and neutrons/antineutrons.

\clearpage
\bibliographystyle{apsrev}
\bibliography{AMON_paper1}

\end{document}